\documentclass[10pt,twocolumn,superscriptaddress,floatfix]{revtex4-2}

\usepackage[utf8]{inputenc}
\usepackage[T1]{fontenc}
\usepackage[colorlinks=true,linkcolor=blue,urlcolor=blue,citecolor=blue,breaklinks=true]{hyperref}

\usepackage{amsmath,amsthm,amssymb,mathtools} 

\usepackage{graphicx}

\usepackage{makecell}
\usepackage{booktabs}
\usepackage{multirow}

\begin{document}
\renewcommand{\acknowledgmentsname}{acknowledgements}

\title{Complexity and Persistence of Price Time
Series\\of the European Electricity Spot Market}

\author{Chengyuan Han}
\affiliation{Forschungszentrum J\"ulich, Institute for Energy and Climate Research - Systems Analysis and Technology Evaluation (IEK-STE), 52428 J\"ulich, Germany}
\affiliation{Institute for Theoretical Physics, University of Cologne, 50937 K\"oln, Germany}

\author{Hannes Hilger}
\affiliation{Institute for Theoretical Physics, University of Cologne, 50937 K\"oln, Germany}

\author{Eva~Mix}
\affiliation{Institute for Theoretical Physics, University of Cologne, 50937 K\"oln, Germany}

\author{Philipp~C.~B{\"o}ttcher}
\affiliation{Forschungszentrum J\"ulich, Institute for Energy and Climate Research - Systems Analysis and Technology Evaluation (IEK-STE), 52428 J\"ulich, Germany}

\author{Mark~Reyers}
\affiliation{Institute for Geophysics and Meteorology, University of Cologne, 50937 K\"oln, Germany}

\author{Christian~Beck}
\affiliation{School of Mathematical Sciences, Queen Mary University of London, London E1 4NS, United Kingdom}
\affiliation{Alan Turing Institute,
London NW1 2DB, United Kingdom}

\author{Dirk~Witthaut}
\affiliation{Forschungszentrum J\"ulich, Institute for Energy and Climate Research - Systems Analysis and Technology Evaluation (IEK-STE), 52428 J\"ulich, Germany}
\affiliation{Institute for Theoretical Physics, University of Cologne, 50937 K\"oln, Germany}

\author{Leonardo~Rydin~Gorj\~ao}
\email[Corresponding author: ]{leonardo.rydin@gmail.com}
\affiliation{Forschungszentrum J\"ulich, Institute for Energy and Climate Research - Systems Analysis and Technology Evaluation (IEK-STE), 52428 J\"ulich, Germany}
\affiliation{Institute for Theoretical Physics, University of Cologne, 50937 K\"oln, Germany}
\affiliation{German Aerospace Center (DLR), Institute of Networked Energy Systems, Oldenburg, Germany}
\affiliation{Department of Computer Science, OsloMet -- Oslo Metropolitan University, N-0130 Oslo, Norway}

\begin{abstract}
    The large variability of renewable power sources is a central challenge in the transition to a sustainable energy system.
    Electricity markets are central for the coordination of electric power generation.
    These markets rely evermore on short-term trading to facilitate the balancing of power generation and demand and to enable systems integration of small producers.
    Electricity prices in these spot markets show pronounced fluctuations, featuring extreme peaks as well as occasional negative prices.
    In this article, we analyse electricity price time series from the European EPEX market, in particular the hourly day-ahead, hourly intraday, and 15-min intraday market prices.
    We quantify the fluctuations, correlations, and extreme events and reveal different time scales in the dynamics of the market.
    The short-term fluctuations show remarkably different characteristics for time scales below and above 12 hours.
    Fluctuations are strongly correlated and persistent below 12 hours, which contributes to extreme price events and a strong multifractal behaviour.
    On longer time scales, they get anti-correlated and price time series revert to their mean, witnessed by a stark decrease of the Hurst coefficient after 12 hours.
    The long-term behaviour is strongly influenced by the evolution of a large-scale weather patterns with a typical time scale of four days.
    We elucidate this dependence in detail using a classification into circulation weather types.
    The separation in time scales enables a superstatistical treatment, which confirms the characteristic time scale of four days, and motivates the use of $q$-Gaussian distributions as the best fit to the empiric distribution of electricity prices.
\end{abstract}

\maketitle

\section{Introduction}
Human reliance on electric power has fostered the development of a large set of technological advances~\cite{USenergy}.
The need to mitigate climate change has, on the one hand, greatly increased the need for low or zero-emission power generation~\cite{REN2021}, and on the other, opened up the electricity markets to small renewable energy power producers~\cite{Green1992,Newbery1998,Woo2011,Cludius2014,Mayer2018,Infield2020}.
Particularly, the short-term markets facilitated the integration of the smaller power producers~\cite{Koch2018}, and have introduced considerable changes to the economic aspects and regulations of electricity markets~\cite{Edenhofer2013,Macedo2020}.
For most power systems, electricity markets are used to trade generated power and guarantee that the consumed power is matched at every point in time~\cite{Rodriguez2014}.
Particularly in open electricity markets -- and for our study here, the European market -- the price of electricity is settled on the Electricity Power Exchange (EPEX SPOT)~\cite{EPEX}.
In the European electricity exchange markets, several different products can be traded, with very different delivery targets and duration.
Particularly on the time scale of days to minutes, there are the day-ahead and intraday markets~\cite{Spodniak2021}, where most renewable energy producers participate.
On the day-ahead market, auction-type products can be traded up to 12 hours before the delivery of power. 
On the intraday markets, continuous-type products are traded up to 5 minutes before delivery~\cite{Narajewski2019,Mix2020}.
Electricity prices are intrinsically coupled in these markets.
The day-ahead markets set an initial value for trading electricity in the intraday market.

From their launch at the beginning of 2010, several studies on the influence, benefits and drawbacks, and impact of short-term electricity markets have been made~\cite{Ocker2017,Mayer2018,Koch2018}.
The need for more accurate forecasting models has also lead to a recent, more close examination of the intraday market prices~\cite{Narajewski2019,Oksuz2019,Janke2019,Marcjasz2020,Abramova2020,Kath2020}.
In this study, we consider a complex systems approach to price dynamics in the purview of stochastic process and examine the interplay between quarter-hourly and hourly price time series, as well as their connection to large weather patterns~\cite{Geman2006}.
We aim at unveiling the characteristic time scales of electricity price time series from a data-driven perspective.
In this, we focus on price dynamics as a complex stochastic process and analyse characteristic metric in the context of complex systems~\cite{Beck1993,Tsallis2009}.
Particularly, we try to answer the questions: what probability distributions do time series show, and to what extent do they differ from Gaussian processes?
What are the reasons for these deviations?
Are price time series stationary processes? If not, do they show distinct time scales? What about stochastic memory and multifractality? Can we quantify those and link them back to how each distinct market works?
And how are they influenced by the weather?
In this work we will address some of these questions, particularly attempting to link the various complex behaviours that price time series exhibit -- such has multifractality and long-range dependence -- to three distinct time scales that we extract from the data.

One should note that electricity prices, unlike most other commodities prices, are distinct in, for example, having not-so-infrequent negative values, being mean-reverting, i.e. returning to some base price after fluctuations, and having pronounced cycles coupled to the generation/consumption of energy, particularly day-night cycles, and other human activities~\cite{Voronin2014,Oksuz2019}.
Being distinctively different from other prices, electricity prices, especially intraday prices, 
have not been extensively analysed from a complex systems' perspective~\cite{Fan2015}. 
An examination of the multifractal properties of electricity price time series is also scarce in the literature~\cite{Norouzzadeh2007,AlvarezRamirez2010}.
We employ the model-free Hilbert--Huang transform (empirical mode decomposition) to remove non-stationarities from price time series~\cite{Huang1998}.
We use multifractal detrended fluctuation analysis~\cite{Peng1994,Peng1995,Kantelhardt2002,RydinGorjao2021} to explain: (i) the time scale separation of the quarter-hourly market below 12 hours and different persistence in the prices; (ii) the anchoring of fast transitions by the day-ahead hourly market at the time scale of $12$ to $48$ hours and the coalescence of precision of all time series~\cite{Wang2013a,Wang2013b}.
We also employ superstatistical methods~\cite{Beck2001,Beck2003,Rak2007} to unveil the longer time scale of prices equilibrium at roughly $96$ hours and obtain the entropic indices of each time series -- a measure of the strength of non-stationarities -- which all differ from $1$~\cite{Gopikrishnan1999,Queiros2005a,Queiros2005b}.
These results support that the statistics of electricity price time series follow a $q$-Gaussian distribution~\cite{Borland2002,Borland2016,Zhao2018}.
The methods employed aim to extract the various mentioned features and time scales solely from the price data, without any other exogenous information.
As a final step, we examine circulation weather types data~~\cite{Jones1993,Reyers2015}, which comprise an objective measure of the state of the flow over Central Europe, particularly describing the strength of the wind.
We show that the prices are inextricably related to large weather parameters, and their statistics change considerably between calmer and strong wind conditions, further justifying our superstatistical approach to price dynamics.

This article is organised as follows: Section~\ref{sec:background} provides a short background on the European electricity markets. Section~\ref{sec:analysis} is comprised of five subsections: Subsection~\ref{sec:stationarity} discusses the aspects of non-stationarity in price time series and how to deal with them; Subsection~\ref{sec:statistics} explains the statistics of price time series and introduces a candidate model to explain these; Subsection~\ref{sec:persistence} discusses simultaneously the intrinsic correlation and persistence in price time series, unveiling our short-term time scale, as well as the rapid jumps in prices, unveiling our mid-term time scale; Subsection~\ref{sec:superstatistics} addresses the change of statistics over time, unveiling our long-term time scale in price time series, and offers a justification for the aforementioned candidate model for price time series statistics; Subsection~\ref{sec:cwt} covers an analysis of the connection between large-scale weather patterns and the changes in statistical properties of the price time series.
Section~\ref{sec:conclusion} provides a set of concluding comments on the results.

\section{Background}\label{sec:background}
A major portion of the Continental European electricity is traded at the European Energy Exchange (EEX).
For the case of Germany and Austria, electricity spot market and over-the counter trading takes place at the European Power Exchange (EPEX SPOT)~\cite{EPEX}, which is a subsidiary of the EEX.
This market is used particularly to balance the daily changes of power in Continental Europe, as well as the very short quarter-hourly and hourly imbalances in power generation and consumption~\cite{MaerkleHuss2018}. 

On futures markets, electricity is often traded weeks, months, or even years before the actual delivery of electricity~\cite{Wilkens2007}.
In contrast to other markets, the supply and demand of electricity has to be met at each point in time to guarantee a stable power system.
While the future demand can be approximated by experience using, for example, the standard load profiles, some deviations might become apparent when getting closer to the date of delivery~\cite{Peters2020}.
Additionally, due to the weather-dependent nature of the increasing share of wind and solar energy resources, it is not possible for a producer to precisely predict the amount of electricity that will be produced at a time in the future~\cite{Borggrefe2011,Behm2020}. 
Thus, shorter-term trading is needed and takes place on the spot markets, making these markets essential instruments for renewable energy source producers.

In Europe, the trading on these shortest time scales is done on the day-ahead and intraday market. 
The time series of three intrinsically connected electricity prices from the spot markets are studied in this article: the day-ahead hourly price time series, the intraday hourly price time series, and the intraday quarter-hourly price time series.
Two distinct markets schemes are present here: (i) the day-ahead or auction market, on which offers can be placed up to 12:00 (noon) prior to the day of effect for the hourly products.
(ii) the intraday or continuous market, on which offers for the subsequent day may be placed from 15:00 (16:00 for quarter-hourly products) of the prior day up to 5 minutes before the respective trading block.
While the last successful bid determines the market clearing price that has to be paid by everyone in the case of the day-ahead market, the intraday market prices are given by a pay-as-bid principle.
In this sense, the intraday market is intrinsically coupled with the day-ahead market, as the day-ahead clearing price serves as a first price reference for the prices in the intraday market.

\begin{figure*}[t]
	\includegraphics[width=\linewidth]{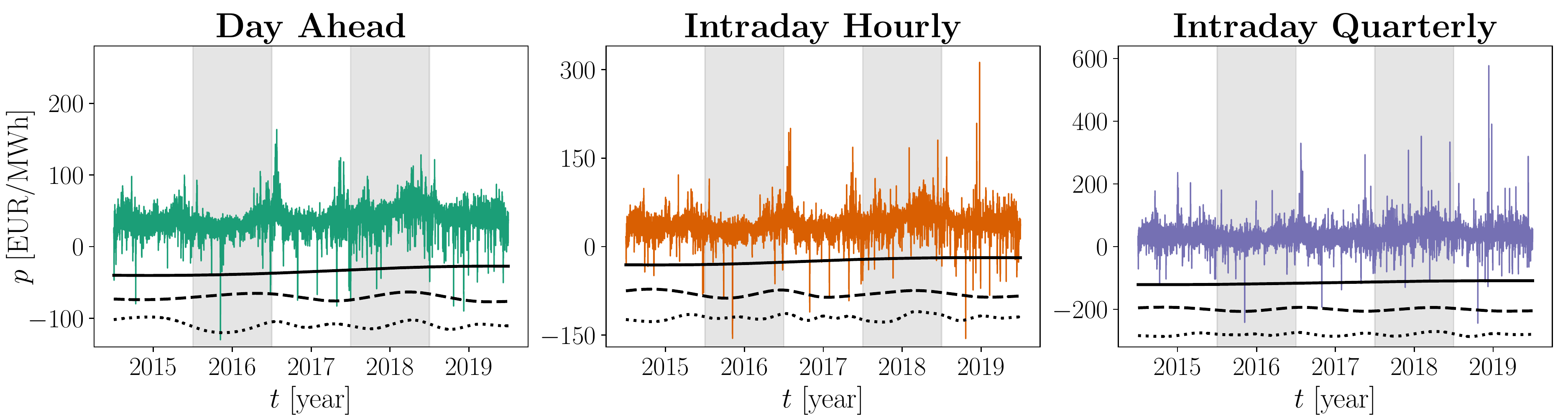}
	\caption{The three electricity price time series examined in this study, from January 2015 to December 2019.
	The average fluctuations around a mean value are visible, as well as occasional jumps into either very large or possibly negative prices.
	The three black curves display the three slowest intrinsic mode functions (IMFs) obtained via the Hilbert--Huang transform, which are subtracted from the data to remove the long-term non-stationarity.
	Data from EPEX, 2015--2019~\cite{EPEX}.
	Figures generated with \texttt{python}'s Matplotlib~\cite{Matplotlib}.}\label{fig:1}
\end{figure*}

Since the trading on the intraday market exists to clear the mismatch that remains after trading on the day-ahead market is finished, it is a smaller market in volume but still an essential one in ensuring the stable operation of the power system.
The volume of trade in 2019 of the day-ahead market totalled $501.6\,$TWh and of the intraday market totalled $83.2\,$TWh~\cite{EPEXAnualReport2019}.

The implementation of the quarter-hourly blocks of trading in 2011 for intraday trading, and extended in 2014 to day-ahead trading, were designed to deal with the increase in renewable energy source input, such as solar and wind energies, which introduces stronger fluctuations in power-grid systems.
Likewise, these short window markets invited various smaller energy producers, particularly of renewable energies, to participate in the market, as they can now trade in time margins wherein they know they can generate the necessary electric power.
Furthermore, the introduction of short trading periods contributes to the improvement of power-grid frequency stability~\cite{Weissbach2008, RydinGorjao2020,Kruse2021}.

\section{Electricity price time series analysis}\label{sec:analysis}
\subsection{Long-term non-stationarity}\label{sec:stationarity}

\begin{figure}[tb]
	\includegraphics[width=\linewidth]{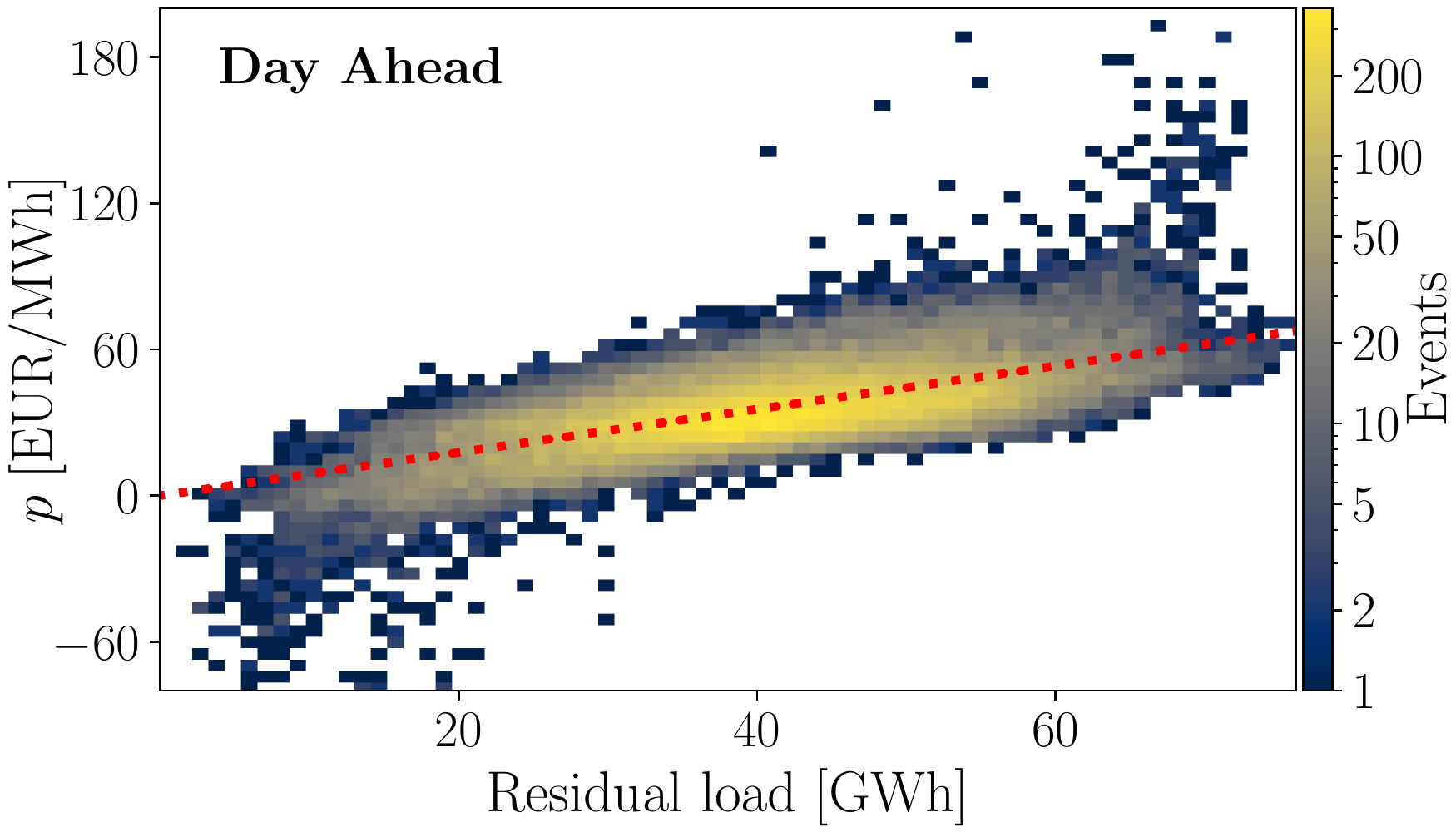}
	\caption{The electricity price strongly depends of the residual load. 
	The figure shows a joint histogram of the price $p(t)$ in the day-ahead hourly market and the residual load, i.e.~the difference of load and renewable generation $D(t) - G_{\mathrm{r}}(t)$ in a colormap plot with a logarithmic scale.
	Assuming a perfect market equilibrium, prices would be given by the function $\bar{p}(t) \approx G_{\mathrm{d}}^{-1}[D(t) - G_{\mathrm{r}}(t)]$, which can be approximated by a linear function (dotted red line). 
	The fluctuations, i.e. deviations from the line, are evident, as well as occasional extreme events.
	Data from EPEX, 2015--2019~\cite{EPEX}.
	}\label{fig:2}
\end{figure}

The dynamics of electricity prices is closely connected to the dynamics of the load and the renewable generation. 
In fact, a rough estimate for the price $p$ at a time $t$ can be obtained from the balance equation of generation and demand, $G_{\mathrm{r}}(t)+G_{\mathrm{d}}(p)=D(t)$.
Here, $G_{\mathrm{d}}(p)$ denotes the supply curve of the dispatchable generation, also called the merit-order curve~\cite{Sensfuss2008,Yang2013,Cludius2014}. 
The demand $D(t)$ and the renewable generation $G_{\mathrm{r}}(t)$ vary strongly in time, whereas the dependence on the price is negligible, i.e. the demand is inelastic. 
Solving for the average price yields $\bar{p}(t) \approx G_{\mathrm{d}}^{-1}[D(t) - G_{\mathrm{r}}(t)]$, i.e. the price dynamics is mainly driven by the load minus the intermittent renewable generation, commonly referred to a the residual load (Fig.~\ref{fig:2}).
The residual load shows pronounced weekly and seasonal pattern and a strong variability on the synoptic scale~\cite{Weber2018}.
It must be kept in mind that this is only a rough estimate, which cannot explain many details, such as the occurrence of negative prices.

Electricity prices in any exchange market are influenced by both short and long-term trends, particularly those reliant on renewable energy sources.
In a single day, electricity prices tend to be lower at night.
The price is also often lower during weekends due to lower consumption~\cite{Braun2018}.
When looking at a longer time period, a more distinct scale emerges: a seasonal and yearly scale~\cite{Halbruegge2021}.
The average price of electricity fluctuates at the level of months, usually culminating in the largest average prices occurring by the end of the year.
These fluctuations make the average of price time series change slowly over the years, e.g., the day-ahead market has seen a variation of the average price from $31.6\,$EUR/MWh in 2015, $29.0\,$EUR/MWh in 2016, $34.2\,$EUR/MWh in 2017, $44.5\,$EUR/MWh in 2018, to $37.7\,$EUR/MWh in 2019.

In this work, we will deal with variations of price time series on different time scales, from scales of $\Delta t < 12$ hours to scales of $\Delta t \!\!\sim\!\!4$ days and longer.
Long-term changes, as those described above, affect the statistics of the time series~\cite{Weron2004b}.
These changes are well understood, yet present a difficult task if we are interested in understanding the fundamental nature and statistics of price time series.
Take the simple example of the variation of the average of the day-ahead price: this implies that examining an aggregated probability distribution of all five years of data will not capture the changes of the average price values that happen yearly.

In order to investigate the short-term variability, long-term trends and periodicities must be separated from the time series. 
To this end, we employ a model-free detrending method to remove the slowest trends in the data.
We use the model-free Hilbert--Huang transform (empirical mode decomposition) method to extract these variations~\cite{Huang1998,Zhaohua2004,Lahmiri2015}.
The Hilbert--Huang transform extracts a set of intrinsic mode functions (IMFs) from the (non-stationary) price time series.
This is achieved via an iterative process of obtaining the set of local minima and maxima of the data and connecting each set via cubic splines, forming an envelope around the time series.
Subsequently, find the middle curve equidistant to the upper (maxima) and lower (minima) envelope.
This is the first IMF.
Subtract this to the actual time series and repeat the procedure to uncover the subsequent IMFs, until the data is solely left with a single residual trend.
One of the main advantage of using the Hilbert--Huang transform is that it can handle non-stationary and non-periodic trends, unlike a Fourier decomposition.

In Fig.~\ref{fig:1}, the full set of five years of data of the three examined time series is displayed, alongside the three slowest IMFs, i.e. the larger wavelength IMFs, which we use to remove the long-term trend.
The data of the day-ahead hourly price, intraday hourly, and intraday quarter-hourly price shows small diffusive-like fluctuations as well as large excursions or jumps.
From hereon, we will work with the detrended data from which the three lowest IMFs have been removed unless stated otherwise.

\subsection{Statistics of electricity price and electricity price increment time series}\label{sec:statistics}
\begin{figure*}[t]
	\includegraphics[width=\linewidth]{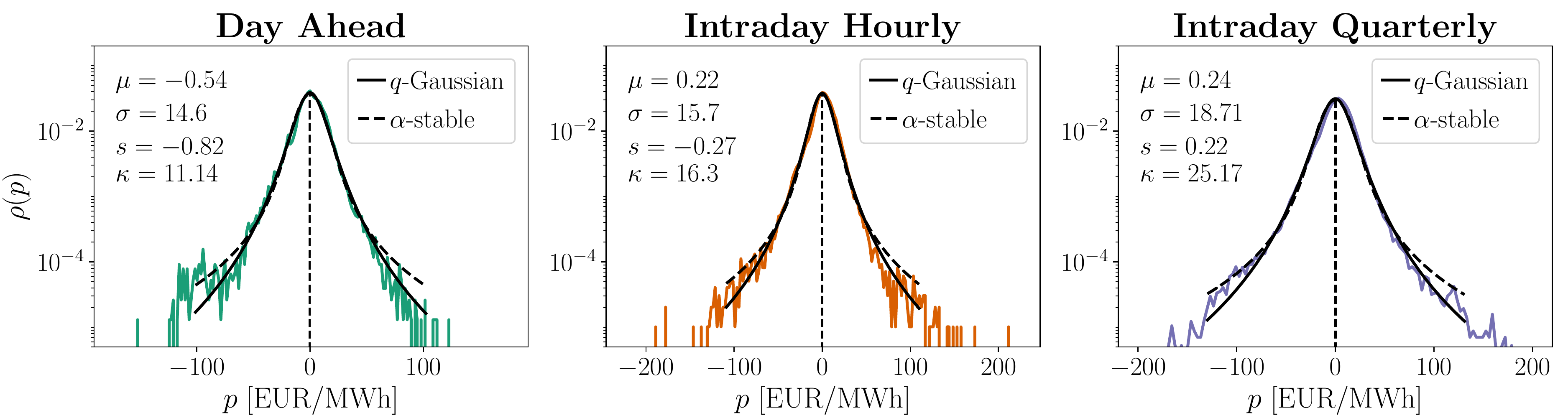}
	\caption{Empirical distributions of the detrended price time series.
	Each empirical distribution is fitted via a maximum likelihood algorithm~\cite{SciPy,NumPy} with a $q$-Gaussian and a symmetric $\alpha$-stable distribution, given by Eq.~\eqref{eq:q_gaussian} and Eq.~\eqref{eq:levy}, respectively.
	The $q$-Gaussian distributions yield a better fit than the $\alpha$-stable distributions.
	The Kullback--Leibler divergence of the empirical and fitted distributions are given in Tab.~\ref{tab:1}, where one can see that the $q$-Gaussian is the best at describing the empiric price time series.
	Mean $\mu$, standard deviation $\sigma$, skewness $s$, and kurtosis $\kappa$ of the empiric data are given in each figure.}\label{fig:3}
\end{figure*}

Having removed the long-term trends of the data, we will now examine the statistics of the data in more detail.
In Fig.~\ref{fig:1}, we observe large excursions of the prices, sometimes even leading to negative prices for all considered time series.
A common method to quantify the dynamics of price time series is to examine the probability distribution or probability density function, as shown in Fig.~\ref{fig:3}. 
One can clearly observe that the data is not described by a Gaussian distribution.
To examine the impact of the heavy tails of the price distributions, we examine the fourth central moment of the price probability distributions, the kurtosis.
The kurtosis of a random variable, or in our case, a price time series $X$ is given by
\begin{equation}\label{eq:kurtosis}
\kappa_X = \mathbb{E}\left[\left(\frac{X - \mu_X}{\sigma_X}\right)^4\right] = \frac{\mathbb{E}\left[(X - \mu_X)^4\right]}{\left(\mathbb{E}\left[(X - \mu_X)^2\right]\right)^2},
\end{equation}
with $\mathbb{E}[\cdot]$ denotes the expected value, $\mu_X$ the mean value of $X$, and $\sigma_X^2$ the variance of $X$.
For example, a Gaussian or normal distribution has a kurtosis of $\kappa_X=3$. 
Any distribution with a kurtosis $\kappa_X>3$ is considered heavy-tailed and is called leptokurtic.
Conversely, if a distribution has a kurtosis $\kappa_X<3$, it is called platykurtic.

The electricity price time series feature pronounced jumps clearly visible in the data (Fig.~\ref{fig:1}). 
Hence, the statistics of the electricity prices can not be expected to be well described by a Gaussian. 
Instead, we expect the distribution to be leptokurtic, i.e. to have a kurtosis $\kappa_X$ larger than $3$.
In Fig.~\ref{fig:3} we display the probability density function $\rho(p)$ of the three detrended price time series (here with a mean close to zero given the detrending performed previously). 
The heavy tails are clearly visible on a semi-logarithmic scale, where a Gaussian distribution would look like an inverted parabola. 
This finding raises the question of what statistics is more suitable to capture the statistical features of electricity prices.

A description of the price time series as a Gaussian process is therefore not suitable, as the empirical distributions of the price time series are highly leptokurtic.
As there is no \textit{a priori} model for these type of data, we will begin by examining the adequacy of describing the data's distribution via two classical distributions: Lévy $\alpha$-stable distributions~\cite{Mandelbrot1960} and $q$-Gaussian distributions~\cite{Tsallis1988}.
These two are chosen for being potentially very leptokurtic distributions, just like the empirical distribution of the data suggests.

The symmetric Lévy $\alpha$-stable distribution has no closed formula for the probability density function $L_{\alpha, c, \mu}(x)$, but it can be expressed via its characteristic function (the Fourier transform of its probability density function) via
\begin{equation}\label{eq:levy}
\begin{aligned}
    L_{\alpha, c, \mu}(x) &= \frac{1}{2\pi}\int\limits_{- \infty}^\infty \varphi_{\alpha, c, \mu}(t)e^{- ixt}\,\mathrm{d}t,~\text{with} \\
    \varphi_{\alpha, c, \mu}(t) &=  \exp \left ( i t \mu - |c t|^\alpha \right ).
\end{aligned}
\end{equation}
Here we will focus only on symmetric (and zero mean $\mu=0$) $\alpha$-stable distributions, but more general asymmetric $\alpha$-stable distributions exist~\cite{Applebaum2011}.

A $q$-Gaussian distribution is a three-parameter distribution with a probability density function $G_{q,c,\mu}(x)$ given by~\cite{Tsallis1988,Umarov2008}
\begin{equation}\label{eq:q_gaussian}
    G_{q,c,\mu}(x) = \frac{\sqrt{c}}{N_q} e_q(-c \left(x-\mu\right)^2),
\end{equation}
where $e_q(\cdot)$ is the $q$-exponential $e_q(\cdot)$ given by
\begin{equation}
    e_q(x) = [1+(1-q)x]^{1/1-q}
\end{equation}
and a normalisation constant $N_q$ given by
\begin{equation}
    N_q = \frac{\sqrt{\pi}~ \Gamma\!\left(\frac{3-q}{2(q-1)}\right)}{\sqrt{q-1}~\Gamma\!\left(\frac{1}{1-q}\right)}, ~\text{for}~ 1 < q < 3,
\end{equation}
where $\Gamma(\cdot)$ is the Gamma function, $\Gamma(n)=(n-1)!$, if $n \in \mathbb{N}$.
Note that these two distribution converge to a normal distribution $\mathcal{N}$ when $q\to1$ or $\alpha\to2$, respectively, i.e. $L_{2,c,0}=G_{1,c,0}=\mathcal{N}(0,1/c)$.
Here we have expressed these two distributions in a comparable way. 
The main interest for us is to ascertain the heavy-tailedness of the distribution, which is described by the value of $q>1$ for the $q$-Gaussian distribution, and the value of $\alpha<2$ for the $\alpha$-stable distribution.
The parameter $c$ is the scale, somewhat related to the variance. 
Note that Lévy $\alpha$-stable distributions do not have a well-defined variance for $\alpha<2$ and $q$-Gaussian distributions have a well-defined variance only if $q<5/3$.
Lastly, $\mu$ is the centre of the distribution, which equals the expected value as long as it exists.

In order to discern which probability distribution function best fits the distribution of our data we evaluate the Kullback--Leibler divergence. 
The Kullback--Leibler divergence $D_{\mathrm{KL}}(r | s)$ of two probability density functions $r$ and $s$, is given by
\begin{equation}\label{eq:KL}
    D_{\mathrm{KL}}(r | s) = \int\limits_{-\infty}^\infty r(x) \ln\left(\frac{r(x)}{s(x)}\right) \mathrm{d}x.
\end{equation}
This, not surprisingly, resembles an entropy formulation, which in this case is the relative entropy of $r$ in relation to $s$.
To be more precise, the Kullback--Leibler divergence $D_{\mathrm{KL}}(r | s)$ should be defined over a set $\mathcal{X}$ with any measure $\mu$ such that $r = \mathrm{d}R/\mathrm{d}\mu$, $s = \mathrm{d}S/\mathrm{d}\mu$, with $R$ and $S$ continuous random variables drawn from $r$ and $s$, respectively.

In Tab.~\ref{tab:1} we show the Kullback--Leibler divergence $D_{\mathrm{KL}}(p|\cdot)$ of the empiric distributions of the three time series in relation to a $\alpha$-stable distribution, as given in Eq.~\eqref{eq:levy}, and to a $q$-Gaussian distribution, as given in Eq.~\eqref{eq:q_gaussian}.
A $q$-Gaussian distribution offers a better fit for all the three price time series.
For each of the fits, we show as well the calculated $\alpha$ and $q$ values for the $\alpha$-stable and $q$-Gaussian distributions, respectively.
The $q$ values will be re-evaluated later in Sec.~\ref{sec:superstatistics} and compared with the entropic indices derived from the data.
It is worth mentioning that for both proposed distributions we observe large deviations from usual Gaussian distributions.

We have thus far concerned ourselves with the statistical properties of the data, detailing a candidate distribution that can capture the heavy tailedness of the leptokurtic distribution of the data.
An equivalently important question relates to the correlations of the time series, in particular, their persistence behaviour, which we address subsequently.

\begin{table}[h]
  \centering
  \begin{tabular}{|r|c|c|c|c|}
    \hline
    \multirow{2}{*}{} & \multicolumn{2}{c|}{$\alpha$-stable}  & \multicolumn{2}{c|}{$q$-Gaussian}\\
    & $D_{\mathrm{KL}}(p|\cdot)$ & $\alpha$ & $D_{\mathrm{KL}}(p|\cdot)$ & $q$ \\ \hline
             Day Ahead & 0.013 & 1.61 &\textbf{0.012} & 1.46\\
       Intraday Hourly & 0.016 & 1.54 &\textbf{0.014} & 1.50 \\
    Intraday Quarterly & 0.012 & 1.61 &\textbf{0.011} & 1.46 \\ \hline
  \end{tabular}
\caption{Kullback--Leibler divergence $D_{\mathrm{KL}}(p|\cdot)$ of the empirical distributions of the three price time series relative to the two candidate distributions: Lévy symmetric $\alpha$-stable distribution $L_{\alpha, c, \mu}$ given in Eq.~\eqref{eq:levy} and the $q$-Gaussian distribution $G_{q,c,\mu}$ given in Eq.~\eqref{eq:q_gaussian}.
The $q$-Gaussian distribution minimises the Kullback--Leibler divergence $D_{\mathrm{KL}}$ for all price time series, indicated by a bold font.
The $\alpha$ and $q$ values of the distributions are given as per the best fit.}\label{tab:1}
\end{table}

\subsection{Persistence and fractality in price time series}\label{sec:persistence}
Separately from time series statistics, the examination of the correlations -- at different temporal scales -- allows us to uncover which phenomena are recurring in a statistical sense.
That is, is the time series persistent and thus repeats itself?
Or is it anti-persistent, and thus follows an opposite tendency in comparison to past events? 
In other words, we are interested in studying the long-term memory or long-range correlations of the data.
When studying stochastic time series, such as price time series, a common method to evaluate the long-range dependency is to estimate the Hurst exponent $H$~\cite{Hurst1951}.
The Hurst exponent $H$ of a time series with uncorrelated increments is $H=0.5$.
One can roughly picture this as imagining that at any point of the time series, the subsequent price is as likely to be higher as it is likely to be lower than the present price.
In this manner, Hurst exponents $H>0.5$ indicate that the increments of the price time series have positive correlations, i.e. are persistent, and thus if we witnessed an increase (decrease) in the price, it is more likely that the price will keep increasing (decreasing).
Conversely, Hurst exponents $H<0.5$ indicate anti-persistence or anti-correlation. Thus a price increase (decrease) is more likely followed by a price decrease (decrease) just after.
This is a vital metric in order to understand whether hedging is possible in electricity markets~\cite{Weron2000}.

A time series can have various Hurst exponents at different scales, telling us, for example, that the price is positively correlated at some very short time scale and negatively correlated at some much larger time scale. 
This, in fact, is what we will see below.
Compounding this, we have also seen in Fig.~\ref{fig:1} the large excursions to very high or negative prices, which we quantified by proposing a suitable candidate distribution for the data.
We will also examine the strength of their fluctuations as we change between time scales in our time series, and in that sense examine the spectrum of multifractality.

A common method to estimate the Hurst exponent, as well as the multifractal spectrum, is Multifractal Detrended Fluctuation Analysis (MFDFA)~\cite{Kantelhardt2002,Peng1994,Ihlen2012,RydinGorjao2021}.
As the name suggest, MFDFA studies the fluctuation of one-dimensional time series around a smooth trend.
First, define the function $F(v,s)$ over the integrated time series $Y_i = \sum_{k=1}^i \left ( X_k - \mu_X \right),~\text{for}~i=1,2, \dots, N$ as
\begin{equation}\label{eq:F(v,s)}
    F(v,r) = \frac{1}{r} \sum_{i=1}^r [Y_{(v-1)r + i} - y_{(v-1)r + i}]^2,
\end{equation}
for $v=1,2, \dots, N_s$. 
Here, $Y_{r}$ is the segmentation of the time series into non-overlapping segments of size $r$, and $y_{r}$ is a polynomial fit to this segment of the data.
It is particularly well adapted to data with trends.
The algorithm first removes the trends of sequential segments of the data by subtracting local polynomial fits to the time series via least-squares and only subsequently calculates the variance of each segment.
We will utilise polynomials of order one.

\begin{figure}[t]
	\includegraphics[width=\linewidth]{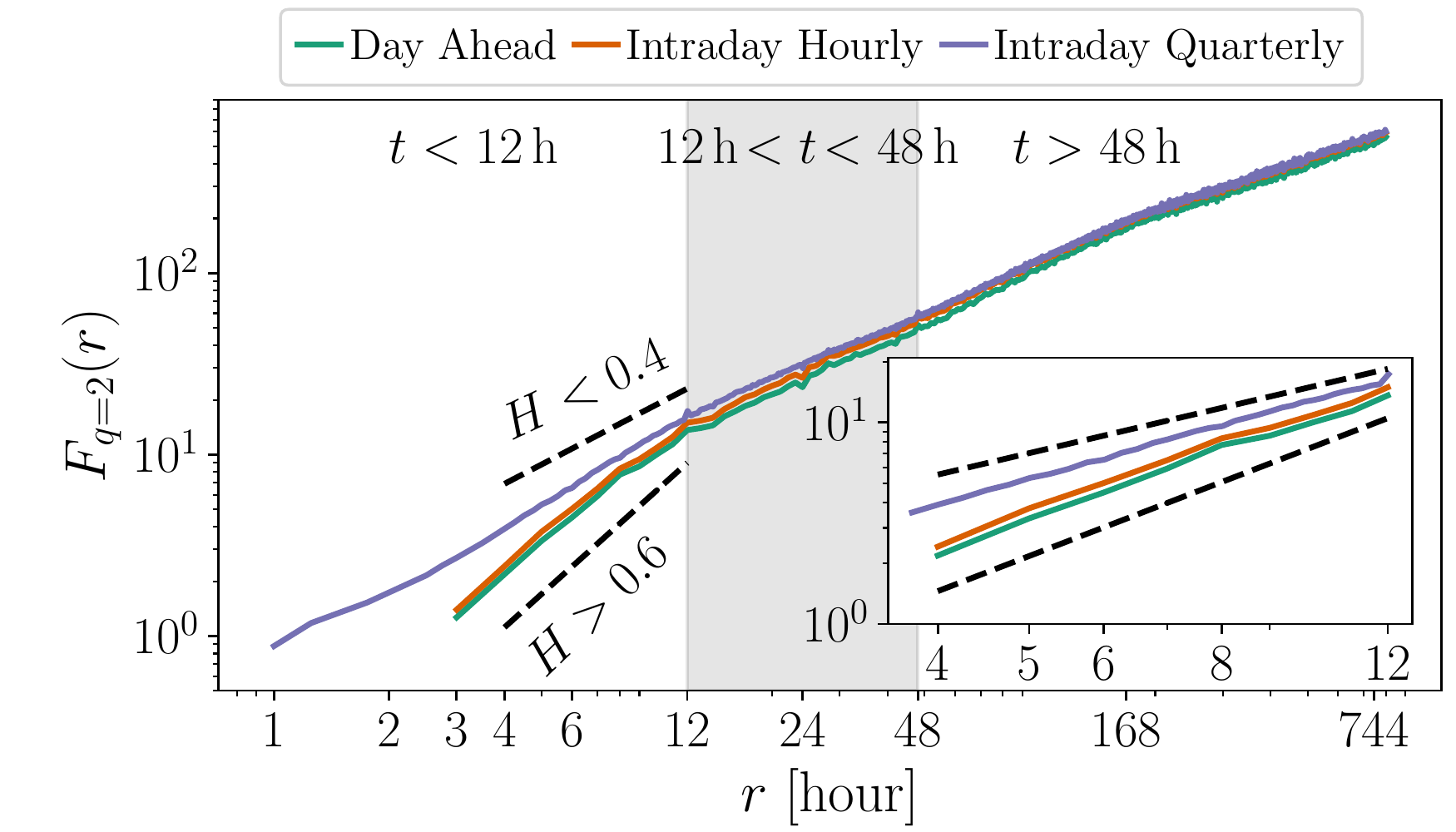}
	\caption{Fluctuation function $F_{\hat{q}=2}(r)$ over the scale $r$ of the three price time series in a double-logarithmic plot.
	By fitting the curves we can extract the Hurst exponent as given in Eq.~\eqref{eq:F_scaling}.
	Noticeable is the change from persistence to anti-persistence at the $12$ hour mark, which is present for the hourly markets but not the quarter-hourly one.
	The separation into three disjoint time ranges of $<12\,$h, $12\,$h$<t<48\,$h, and $t>48\,$h is discussed in the multifractal analysis in Fig.~\ref{fig:5}.}\label{fig:4}
\end{figure}

Subsequently, to extract the multifractal spectrum of the time series, a set of different powers are taken over the average of all segments.
From this, we define the fluctuation function $F_{\hat{q}}(s)$, which depends on a time scale $s$ of the process and the aforementioned powers $\hat{q}$
\begin{equation}\label{eq:F}
  F_{\hat{q}}(r) = \left\{\frac{1}{N_r} \sum_{v=1}^{N_r} [F(v,r)]^{\hat{q}/2}\right\}^{1/\hat{q}}.
\end{equation}
This is similar to considering a set of equivalent norms with different powers, e.g., $L_1, L_2, L_3,$ etc.
This is the function we will study here onward.
We note that the index $\hat{q}$ is not related to the $q$-parameter in the distribution in \eqref{eq:q_gaussian}.
The fluctuation function $F_{\hat{q}}(r)$ captures the increase of the variance of the segments of the time series, i.e.
\begin{equation}\label{eq:F_scaling}
  F_{\hat{q}}(r) \sim r^{h(\hat{q})}
\end{equation}
where $h(\hat{q})$ is known as the generalised Hurst exponent, which for the case of $\hat{q}=2$ reduces to our aforementioned Hurst exponent $H=h(\hat{q}=2)$.

We expect the price time series are not monofractal processes, i.e. processes that are quantified by solely a single Hurst exponent $H$.
In order to quantify the influence of the jumps in the time series, we turn to the dependence of $h(\hat{q})$ on $\hat{q}$.
From this, we can construct the singularity spectrum $f(\alpha)$ and the singularity strength $\alpha$ as the Legendre transform given by
\begin{equation}\label{eq:sing_strength}
    \hat{\alpha} = h(\hat{q}) + \hat{q}h'(\hat{q}),
\end{equation}
and
\begin{equation}\label{eq:sing_spect}
    f(\hat{\alpha}) = \hat{q}[\hat{\alpha} - h(\hat{q})] + 1,
\end{equation}
with $h'(\hat{q})=\mathrm{d}h(\hat{q})/\mathrm{d}\hat{q}$.
This leads to the iconic shape of the singularity spectrum $f(\hat{\alpha})$ as one half of an inverted parabola with maximum at $\hat{\alpha}=0$~\cite{Ihlen2012}.
Similarly as before, we note that singularity strength $\hat{\alpha}$ is not related to the $\alpha$ for $\alpha$-stable distributions in \eqref{eq:levy}.

\paragraph{Persistence of price time series}
First, we turn to the question of long-range dependence, i.e. persistence.
In Fig.~\ref{fig:4}, we show the fluctuation function $F_{\hat{q}=2}(r)$ for $\hat{q}=2$.
We plot $F_{\hat{q}=2}(r)$ versus the scale $r$ in a double logarithmic scale, as we are interested in the exponents of Eq.~\eqref{eq:F_scaling}, i.e. the Hurst exponent $H$.
The exponent, in a double-logarithmic plot, is simply the slope of the curves, which we extract by fitting a straight line.
Immediately, two phenomena are striking: First of all, for time scales larger than $12$ hours, all time series have virtually identical anti-correlations, with $H\approx 0.16$.
Moreover, at time scales smaller than $12$ hours, the larger hourly markets become positively correlated, with $H\approx0.63$ for the day-ahead hourly price and $H\approx0.61$ for intraday hourly price, whereas the intraday quarter-hourly price does not show a change from correlated ($r<12\,$h) to anti-correlated ($r>12\,$h) behaviour, having $H\approx0.31$.

\paragraph{Fractality of price time series}
We have already mentioned the necessity to properly quantify the effects of the jumps in the time series, and we have introduced the singularity spectrum $f(\alpha)$.
Moreover, already in Fig.~\ref{fig:4}, we have found at least two separate time scales for the hourly markets: $<12\,$h and $>12\,$h.
We will now further divide the larger time scale again into periods of $12\,$h$<t<48\,$h and periods larger than $t>48\,$h and study these three time scales and their multifractal spectrum.

\begin{figure}[t]
	\includegraphics[width=\linewidth]{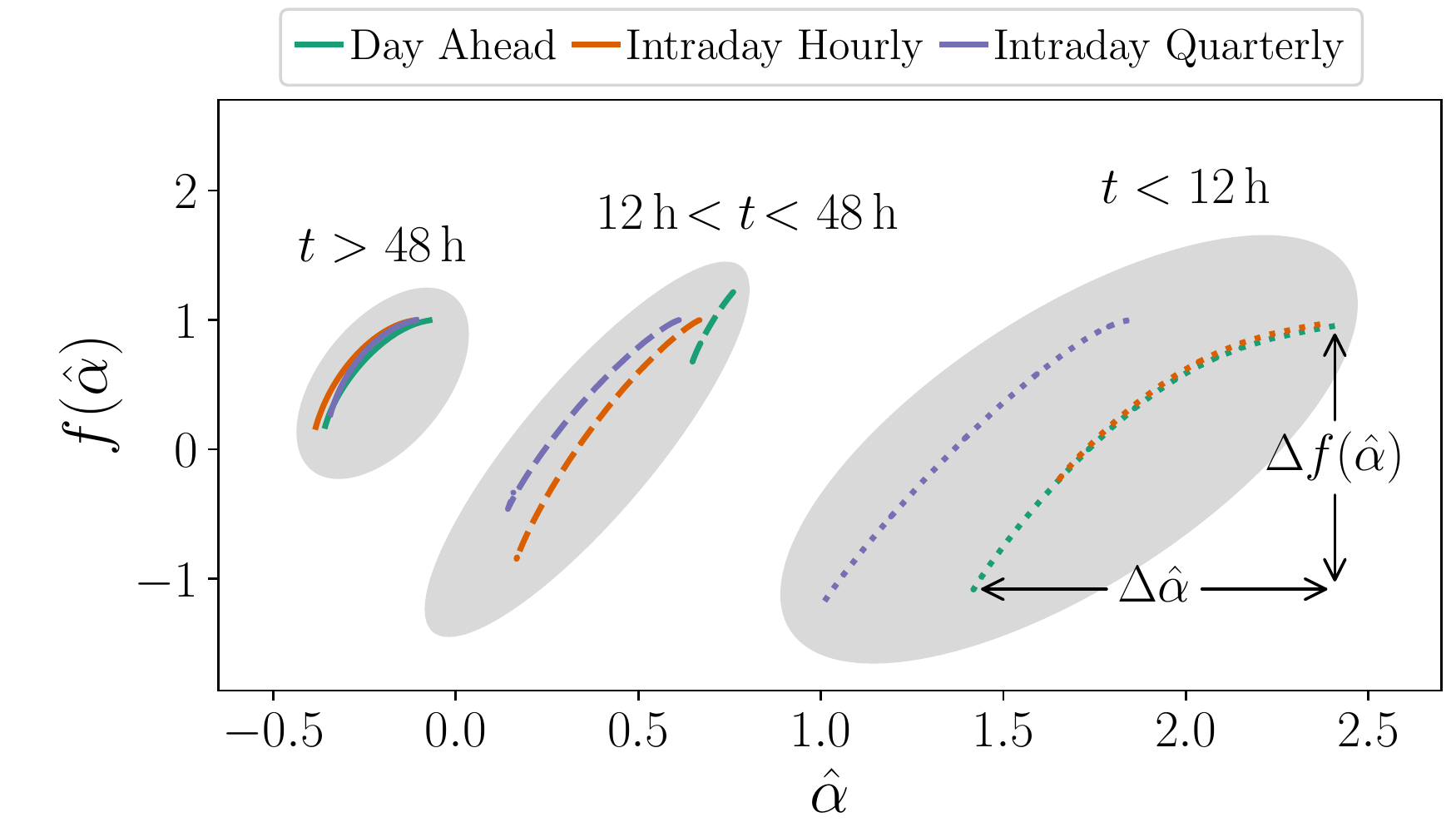}
	\caption{Singularity spectrum $f(\hat{\alpha})$ and singularity strength $\hat{\alpha}$ for three distinct time scales of the three price time series: $t<12\,$h, $12\,$h$<t<48\,$h, and $t>48\,$h.
	The horizontal width, i.e. the multifractal spectrum width $\Delta \hat{\alpha}$, indicates the strength of the fluctuation and jumps at the indicated time scales.
	The data shows larger $\Delta \hat{\alpha}$ at short time scales, indicating that within these windows, very large price variations are seen and are reverted back to their mean value.
	The effects become milder as the time scales increase, telling us that the market shows weaker jumps at large time scales and always reverts back to its mean value.
	Finally, at scales $t>48\,$h, the three markets become indistinguishable, as we had also seen in Fig.~\ref{fig:4} at the same scales.
	Results in Tab.~\ref{tab:2}}\label{fig:5}
\end{figure}

In Fig.~\ref{fig:5}, we display the singularity spectrum for $\hat{q}\in(0,10]$, i.e. the positive half of $f(\alpha)$ for the three aforementioned time scales.
We cannot evaluate the negative $\hat{q}$ powers here due to the limited precision of the data.
To evaluate the meaning of the singularity spectrum, we focus on the widths of $\hat{\alpha}$ for the different time series, i.e.
\begin{equation}\label{eq:delta_h}
    \Delta \hat{\alpha} = \operatorname{argmax}(f(\hat{\alpha})) - \mathrm{min}(\hat{\alpha}).
\end{equation}
This yields a measure of ``how many fractal scales'' are present in each time series, which is commonly denoted as multifractal spectrum width.
If our time series had a single scale, i.e. a single Hurst exponent $H$, then $\Delta \hat{\alpha} = 0$, and we would classify it as monofractal.
The meaning of $\Delta \hat{\alpha}$ is thus straightforward to understand.
If the data is multifractal, i.e. it shows a range of small and large fluctuations and jumps, then $\Delta \hat{\alpha}>0$.
In Tab.~\ref{tab:2} we report all $\Delta \hat{\alpha}$, as given in Eq.~\eqref{eq:delta_h}, as well as $\Delta f(\hat{\alpha}) = \mathrm{max}(f(\hat{\alpha})) - \mathrm{min}(f(\hat{\alpha}))$.
These values give us a notion of which scales show rougher behaviours and which are milder. 

Again here, as before in Fig.~\ref{fig:4}, at large scales, the time series coalesce to having identical fractal behaviour and comparably small $\Delta \hat{\alpha} \approx 0.50$. 
All $F_{\hat{q}=2}$ curves overlay for $t>48\,$h, i.e. all have the same Hurst index $H$.
They all similarly show the same multifractal spectrum.
The very short time scales of $t<12\,$h show the largest $\Delta \hat{\alpha}$,
indicating the strongest multifractal behaviour.
This is very much in line with the rare, sudden price increases or decreases to extreme values, which very quickly correct themselves and return to their average price.

Interestingly, there are considerable differences between the markets in the range $12\,$h$<t<48\,$h. The day-ahead hourly market shows the smallest $\Delta \hat{\alpha}$, i.e.~it shows the weakest multifractality. 
The day-ahead market constitutes the largest share of the electricity markets in volume, and thus ensures that electricity prices must all coalesce to the mean behaviour within the time scale $12\,$h$\,\sim\!48\,$h.
It is therefore likely that, due to the large volume of trade in the day-ahead market and the small $\Delta \hat{\alpha}$ in the range $12\,$h$<t<48\,$h, the day-ahead market serves as an anchor for the other smaller markets and their prices, guaranteeing that, in the long run,
very large prices fluctuations return to normal price ranges within time scales smaller than $48\,$h. 

\begin{table}[h]
  \centering
  \begin{tabular}{|r|c|c|c|c|c|c|}
    \hline
    \multirow{2}{*}{} & \multicolumn{2}{c|}{$>48\,$h}  & \multicolumn{2}{c|}{$12\,\mathrm{h}<t<48\,$h} & \multicolumn{2}{c|}{$t<12\,$h}\\
    & $\Delta \hat{\alpha}$ & $\Delta f(\hat{\alpha})$ & $\Delta \hat{\alpha}$ & $\Delta f(\hat{\alpha})$ & $\Delta \hat{\alpha}$ & $\Delta f(\hat{\alpha})$ \\ \hline
    Day Ahead & 0.29 & 0.82 & 0.14 & 0.59 & 0.99 & 2.05 \\
    Intraday Hourly & 0.27 & 0.82 & 0.50 & 1.35 & 0.71 & 1.20 \\
    Intraday Quarterly & 0.24 & 0.72 & 0.47 & 1.46 & 0.84 & 2.21 \\ \hline
  \end{tabular}
\caption{Multifractal spectrum width $\Delta \hat{\alpha}$ and $\Delta f(\hat{\alpha})$ for three distinct time scales of the three price time series: $t<12\,$h, $12\,$h$<t<48\,$h, and $t>48\,$h.
The smallest time scales show the largest values of $\Delta \hat{\alpha}$.
Overall, the day-ahead hourly market shows the smallest multifractal spectrum width $\Delta \hat{\alpha}$ at the time scales $12\,\mathrm{h}<t<48\,$h, which we propose offers a kind of ``anchor'' for the prices to coalesce around.}\label{tab:2}
\end{table}

\subsection{Obtaining local equilibria in leptokurtic electricity price time series}\label{sec:superstatistics}

We have thus far given an account of the correlation behaviour in the three electricity price time series, unveiling different persistence behaviour between the hourly markets and the quarter-hourly one.
We have also seen that roughly at scales larger than $t>48$ hours, the markets coalesce to a single identical behaviour, both in their diffusive behaviour, seen in Fig.~\ref{fig:4} as well as in their multifractal behaviour, e.g., the presence of jumps in the data, as seen in Fig.~\ref{fig:5}.
Given that all price time series eventually return to an average price value, and that intrinsic periods are present in the data, a time scale at which an equilibrium is reached must exist.
This is the time the price statistics balances out before it is again affected by its various intrinsic changes and large price variations.

\begin{figure}[t]
	\includegraphics[width=\linewidth]{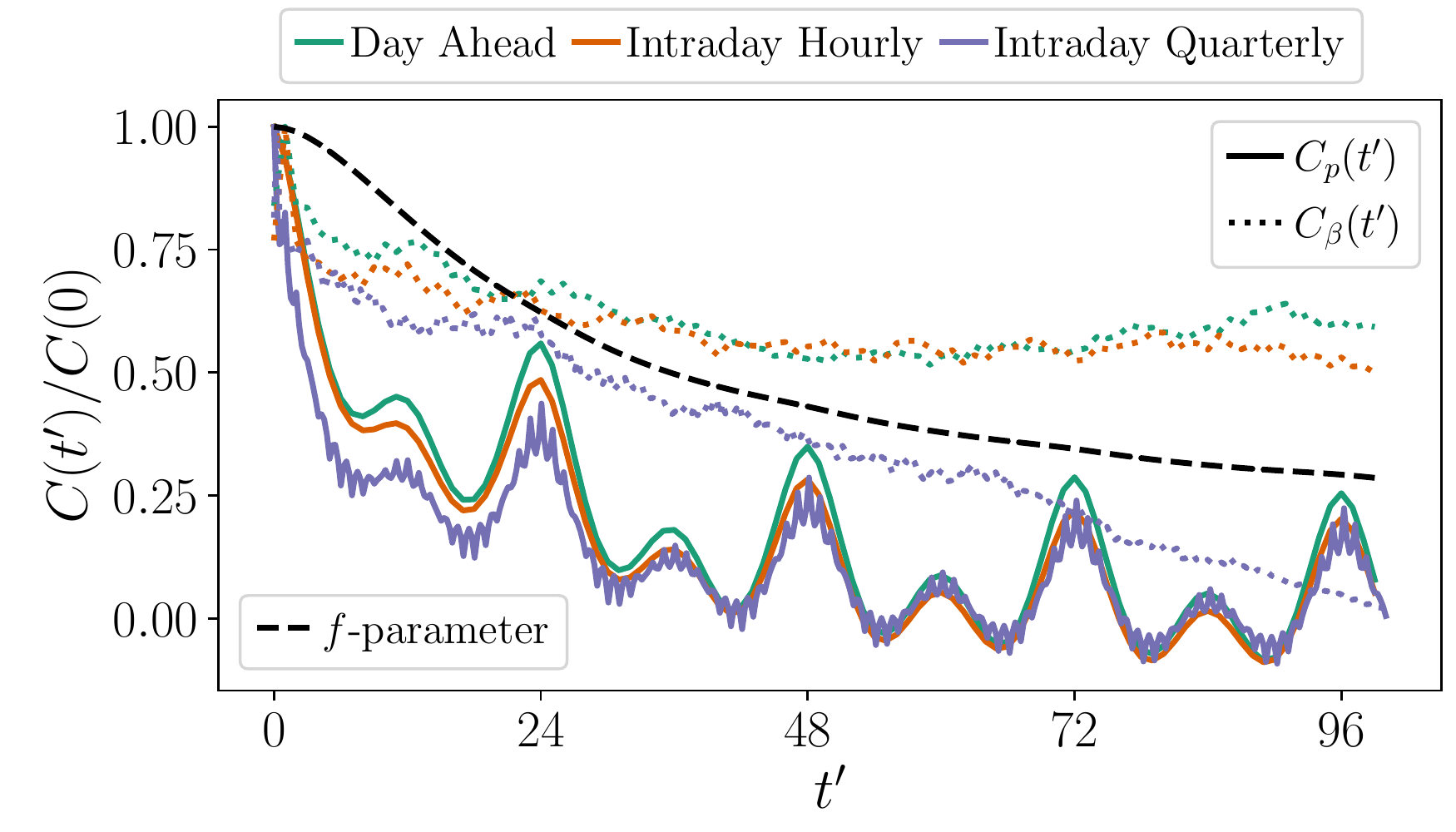}
	\caption{Auto-correlation functions $C(t')/C(0)$ of the price time series and their respective volatilities. 
	The short superstatistical times $\tau$ for each price time series is extracted from the initial exponential decay of $C_p(t')$. 
	The results are given in Tab.~\ref{tab:3}.
	Shown as well is the auto-correlation of the $f$-parameter, discussed in Sec.~\ref{sec:cwt}.}\label{fig:6}
\end{figure}

The most straightforward way to examine the typical time scale of local relaxation of a time series of a stochastic process is to study its auto-correlation function.
The auto-correlation of a time series is given by
\begin{equation}
    C(t-t') = \mathbb{E}[(X(t)-\mu_X)(X(t')-\mu_X)].
\end{equation}
For $t-t'=0$, i.e. $C(0)=\sigma_X^2$ we recover the variance of the process.
For $t\neq t'$ we obtain the covariance, which yields the self-correlation of the process with itself, that is, its memory.
In Fig.~\ref{fig:6}, we show the auto-correlation function $C(t')/C(0)$ for our three price time series.
Along with an exponential-like decay, one can find well defined peaks that indicate the usual periods known in these time series: $12$ hours, due to the day-night cycle, $24$ hours, $48$ hours, etc.
Although there are several peaks, the auto-correlation shows a decay which has a minimum at roughly $90$ hours.
This indicates that this is the time scale at which the process loses its memory.
In order to more precisely ascertain what the intrinsic time scale is for which the price time series attain a local equilibrium, we turn to a superstatistical description~\cite{Beck2001,Beck2003}.

We have already seen that the probability distribution functions of each price time series had very large kurtosis, i.e. they are leptokurtic distributions (see Fig.~\ref{fig:3}).
We further identified the $q$-Gaussian distribution in Eq.~\eqref{eq:q_gaussian} as a good candidate to model the large kurtosis.
Along similar lines as discussed above, we now propose that the underlying stochastic process that gives rise to these complicated distributions of price time series is a composite process with a probability density function $\rho(p)$ given by
\begin{equation}\label{eq:fund_super_dist}
    \rho(p) = \int\limits_{0}^{\infty} f(\beta) \Pi(p|\beta) \mathrm{d}\beta,
\end{equation}
where $\Pi(p|\beta)$ is a conditional distribution dependent on a volatility parameter $\beta$ which has the probability density function $f(\beta)$.

What we are considering here is that price time series consists of two processes: 
One fast process of the actual price time series, with local temporal properties, i.e. a certain level of fluctuations and an average price, and another far slower process, which changes the strength of the fluctuations and/or the average price at a larger temporal scale.
The description above in Eq.~\eqref{eq:fund_super_dist} accounts for the distribution of the price time series $\rho(p)$ being a convolution of these two processes.

Stemming from a physical understanding of the price of stocks in other markets outside power systems~\cite{Borland2002,Borland2002b,vanderStraeten2009}, a common assumption is to take $\Pi(p|\beta)$ as a Gaussian distribution.
This assumption would mean that fundamentally price time series obey locally Gaussian statistics, which is then affected by a superstatistical change given by $f(\beta)$.
In this work, we will relax this constraint and propose that $\Pi(p|\beta)$ need not necessarily be a Gaussian distribution, but instead, simply restrict $\Pi(p|\beta)$ to be a symmetric distribution.
This proposal means that in principle $\Pi(p|\beta)$ can be, for example, a symmetric $\alpha$-stable distribution, or a $q$-Gaussian, or possibly another symmetric distributions (or even just a Gaussian distribution).

To evaluate if a distribution is symmetric, one can evaluate its skewness $s$, i.e. 
\begin{equation}\label{eq:skewness}
s_X \coloneqq \mathbb{E}\left[\left(\frac{X - \mu_X}{\sigma_X}\right)^3\right] = \frac{\mathbb{E}\left[(X - \mu_X)^3\right]}{\left(\mathbb{E}\left[(X - \mu_X)^2\right]\right)^{3/2}},
\end{equation}
which is vanishing if the distribution is symmetric.
As mentioned before, we are interested in finding an average time scale at which the price attains equilibrium.
By this, we mean that we are interested in a point in time where, on average, a segment of the price time series has a skewness $s=0$.
This point tells us, statistically, that the distribution of events around a local price average balances to be symmetrically distributed.
So, in some sense, this is the point in time where the markets average out their electricity price, and they are as likely to see a subsequent increase as a subsequent decrease of the price, statistically speaking.
This time is referred to as the long superstatistical time $T$.
We can estimate the long superstatistical time $T$ by taking segments of the price time series with a given time range $\delta t$
\begin{equation}\label{eq:local_skewness}
    s_p(\delta t) = \left\langle \frac{\frac{1}{\delta t}\sum_{i=(j-1)\delta t +1}^{j\delta t} ({p}_i - \bar{p}_i)^3)}{(\frac{1}{\delta t}\sum_{i=(j-1)\delta t +1}^{j\delta t} ({p}_i - \bar{p}_i)^2)^{3/2}} \right\rangle_{\delta t}
\end{equation}
where $T$ is the defined as the particular $\delta t$ value such that $s(\delta t = T)=0$. 
Previous methods used the kurtosis $\kappa$ rather than the skewness $s$ to estimate the long superstatistical time $T$, but we think that for electricity prices the skewness is a particularly well suited observable, given that electricity prices show both small and large deviations to high or low (and negative) prices at different points in time. 
At $\delta t = T$ both positive and negative tails are equally pronounced, indicating a symmetry of high and low extremes.

\begin{figure}[t]
	\includegraphics[width=\linewidth]{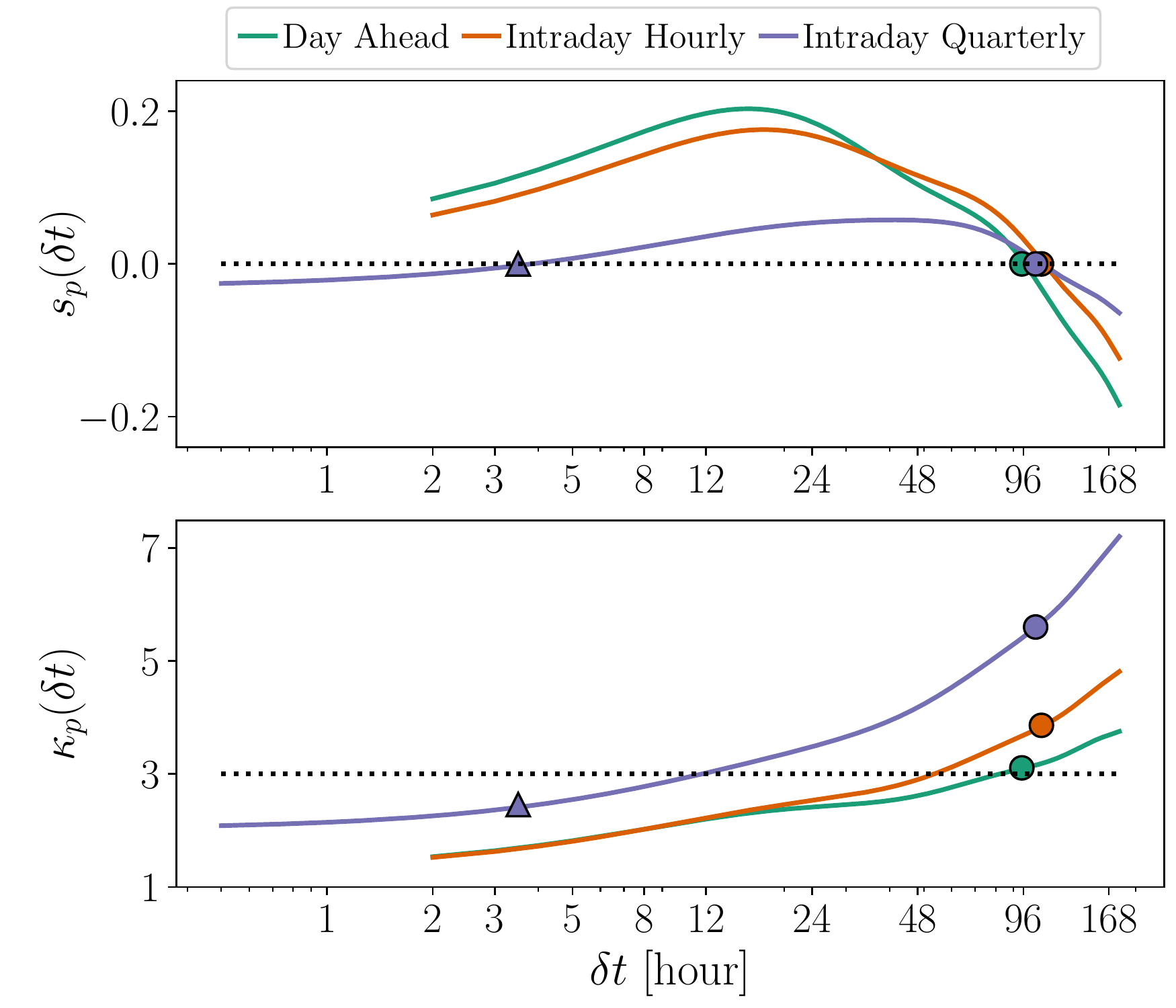}
	\caption{Estimating the long superstatistical time $T$ from the vanishing of the local skewness $s(\delta t)$ given in Eq.~\eqref{eq:local_skewness}.
	The top panel shows the $s(\delta t)$ for increasing segments of time length $\delta t$.
	At approximately $4$ days, indicated with the circular markers, all price time series show a vanishing local skewness.
	We define this as marking the long superstatistical time $T$.
	The lower panel shows the local kurtosis $\kappa_p(\delta t)$.
	Indicated as well is the kurtosis of a Gaussian distribution, i.e. $\kappa=3$, for comparison.
	We see that the time series attain an equilibrium at the vanishing skewness $s$, but they do so with different kurtosis, implying that their equilibrium distributions are not Gaussian distributions, possibly apart from the day-ahead hourly price time series. 
	Each long superstatistical time $T$ can be found in Tab.~\ref{tab:3}.}\label{fig:7}
\end{figure}

In Fig.~\ref{fig:7}, we display the local skewness $s_p(\delta t)$ (top panel) as a function of the time range $\delta t$.
For comparison, we also show the local kurtosis $\kappa_p(\delta t)$ (bottom panel), calculated similarly as in Eq.~\eqref{eq:local_skewness} but considering the kurtosis $\kappa$ as in Eq.~\eqref{eq:kurtosis}.
Interestingly, we see that all three electricity markets attain a skewness of $s=0$, i.e. become symmetrical, at a scale of roughly $4$ days or $96$ hours.
In Fig.~\ref{fig:7}, top panel, we indicate these times with the circular markers.
For comparison, we indicate as well, on the bottom panel, the long superstatistical time and the kurtosis of each of the local distributions.
The dotted line indicates the kurtosis $\kappa = 3$ of a Gaussian distribution.
Although the larger, day-ahead market has a kurtosis very close to that of a Gaussian distribution, the other markets deviate from this and show a large kurtosis $\kappa>3$.
We also note a second transition in the intraday quarter-hourly market at roughly $4$ hours, indicated with a triangular marker, which is yet another point with zero skewness.

Thus, we have found the long superstatistical time $T$ for the three price time series by assuming that these can have a rather general distribution locally, for as long as it is symmetric.
All the markets seem to have a very similar long superstatistical time $T$, pointing to this being the local balancing value for all time series, i.e. this is a common feature of all the markets, likely due to their coupled structure.
The exact long superstatistical times $T$ for each price time series can be found in Tab.~\ref{tab:3}.

From this point, we can proceed further and analyse the volatilities $\beta$ that give rise to the superstatistical distributions for the given time series.
Having unveiled the long superstatistical time $T$ of each time series, we can study the stochastic process of volatilities $\beta$, which is given by,
\begin{equation}\label{eq:beta}
    \beta(t) = \frac{1}{\langle p^2 \rangle_{T} - \langle p \rangle_{T}^2},
\end{equation}
i.e. it is given by the inverse of the local variance of the segments with a time length of $T$.
Strictly speaking, $\beta^{-1}$ is the volatility, as it is proportional to the variance, but for brevity, we will simply denote $\beta$ as volatility.
We can picture this in a simple way: if no changes were happening at a larger time scale (at the long superstatistical time scale) in our time series, then the variance of each segment of time length $T$ would be the same, and thus the volatility $\beta$ would be a constant $\beta_0$.
This would also mean that in Eq.~\eqref{eq:fund_super_dist} the distribution $f(\beta)$ of the volatility $\beta$ would by a delta Dirac distribution and there would be no superstatistical change in the time series ($\rho(p)\equiv \Pi(p|\beta_0)$).
The distributions of the volatilities for our data can be found in App.~\ref{app:1}.

Before we proceed, we need to ensure our superstatistical approach is justified.
Just as discussed before, is it true that we can separate two time scales from the time series? 
For our superstatistical description of the leptokurtic distributions of price time series to be justified, we need to evaluate the correlations of both the time series themselves and of the volatilities.
We need to evaluate what is the typical correlation length of the time series $p(t)$.
This we can do by considering the initial exponential decay of the auto-correlation function of the price time series, such that $C(\tau) = e^{-1}C(0)$, as given in Fig.~\ref{fig:6}.
For superstatistics to be justified, the correlation time $\tau$, denoting the short superstatistical time scale $\tau$, needs to be smaller than the long superstatistical time $T$.
This restriction ensures that a local equilibrium is achieved on a time scale shorter than the long superstatistical time scale $T$.
Fig.~\ref{fig:6} shows the auto-correlation functions, i.e. $C(t')/C(0)$, for the three price time series $p(t)$ and their respective volatilities $\beta(t)$.
In Tab.~\ref{tab:3} we can see that the short superstatistical times $\tau$ are all smaller than the long superstatistical time $T$.
In a similar manner, we can see that the auto-correlation functions of the volatilities $C_\beta(t')$ decay slower than the normalised auto-correlation functions of their respective price time series, telling us the superstatistical changes happen slower than the changes in the time series themselves, as required by the superstatistical modelling approach.

We have thus far shown that our description of the probability density function as a superposition of symmetric (yet unspecified) distributions is justified and seem to indicate that all price time series attain an equilibrium after roughly $4$ days ($96$ hours).
We will now evaluate the strength of the changes of the volatilities $\beta$.
Since we assumed a general description of $\Pi(p|\beta)$ as simply being a symmetric distribution, and given that we have not detailed specifically the distribution $f(\beta)$ of the volatilities $\beta$, we cannot evaluate Eq.~\eqref{eq:fund_super_dist} explicitly.
We can nevertheless consider the integration in Eq.~\eqref{eq:fund_super_dist} for small fluctuations of $\beta$ around $\beta_0=\langle  \beta \rangle$.
For small variance $\sigma^2 = \langle  \beta^2 \rangle - \beta_0^2$ we obtain
\begin{equation}\label{eq:approx-SS}
\begin{aligned}
    \rho(p) &=  \langle  \Pi(p|\beta)  \rangle = \Pi(p|\beta_0) \langle  \Pi(p|(\beta-\beta_0)  \rangle\\
    & =\Pi(p|\beta_0)\left(1 + \frac{1}{2}\sigma^2 p^2 +  \mathcal{O}(\sigma^3)\right) \nonumber \\
    &= \Pi(p|\beta_0)\left(1 + \frac{1}{2} (\bar{q}-1)\beta_0^2 p^2 + \mathcal{O}(\sigma^3)\right),
\end{aligned}
\end{equation}
where we introduce the entropic index $\bar{q}$~\cite{Beck2003,Beck2009}
\begin{equation}\label{eq:q}
\bar{q}=\frac{\langle \beta^2 \rangle}{\langle \beta \rangle^2}.
\end{equation}
The entropic index $\bar{q}$ accounts for the variations of the volatilites.
It is a rather elegant measure for the existence of non-stationarities or non-extensive properties in the data.
As described before, if there were no changes in the variance of the price time series, $\bar{q}=1$, and Eq.~\eqref{eq:approx-SS} would collapse to the case where $\rho(p)\equiv \Pi(p|\beta_0)$.
If, on the other hand, entropic index $\bar{q}$ differs from $1$, then we necessarily have some variation of the volatilities $\beta$.
In Tab.~\ref{tab:3} we report the entropic indices $\bar{q}$ of the three price time series, which all differ strongly from $1$.

\begin{table}[h]
  \centering
  \begin{tabular}{|r|c|c|c|}
  \hline
    & $T$ & $\tau$ &$\bar{q}$ \\\hline
     Day Ahead & $95\,$h & $13.5\,$h & $1.55\,$ \\
    Intraday Hourly & $108\,$h & $11.8\,$h & $1.61\,$ \\
    Intraday Quarterly & $104\,$h & $7.6\,$h & $1.46\,$ \\ \hline
  \end{tabular}
\caption{Long and short superstatistical times $T$ and $\tau$, and the entropic indices $\bar{q}$ of the three price time series. 
In all cases the short superstatistical time is substantially smaller than the long superstatistical time, i.e. $\tau\ll T$.}\label{tab:3}
\end{table}

Naturally a subsequent question is related to our choice of fitting $q$-Gaussian distributions to $\rho(p)$ in Sec.~\ref{sec:statistics}, seen in Fig.~\ref{fig:3}, and the relation of that fitting parameter $q$ with the entropic index $\bar{q}$.
One way a $q$-Gaussian arises is if choose $\Pi(p|\beta)$ as a Gaussian distribution and $f(\beta)$ a Gamma distribution.
From this choice, one finds $q=\bar{q}$.
From a theoretical point-of-view, without detailing the distribution of the volatilities $f(\beta)$, it is not possible to ascertain if $\rho(p)$ is justifiably given by a $q$-Gaussian distribution.
This nevertheless does not preclude comparing the entropic indices $\bar{q}$ with the $q$-values of the best fitting $q$-Gaussian distributions.

In Tab.~\ref{tab:3}, we indicate the entropic indices $\bar{q}$ of the volatilities of each price time series.
We see that these very closely resemble the $q$-values of the $q$-Gaussian distributions in Tab.~\ref{tab:1}.
We have for the day-ahead hourly price time series $q=1.46$ and $\bar{q}=1.55$; for the intraday hourly price time series $q=1.50$ and $\bar{q}=1.61$; and for the intraday quarter-hourly price time series $q=1.46$ and $\bar{q}=1.46$.
These stark similarities offer a justification for the choice of $q$-Gaussian distribution as the descriptors for the distribution of price time series. 
Note that for all three electricity price time series the extracted $\bar{q}$ values are considerably bigger than for other financial time series, such as, e.g., share price indices or foreign currency exchange rates~\cite{Beck2009}.
This is understandable, given the complexity of the demand dynamics of electricity markets.

\subsection{The impact of weather on electricity prices}\label{sec:cwt}

Fluctuations in renewable energy production on different time scales are strongly influenced by weather regimes and systems, like e.g., blocking regimes, low pressure systems, and the passage of fronts~\cite{Wiel2019,Grams2017,Dalton2019}.
Inherently, so are electricity prices because of the \textit{merit-order} effect. 
As previously shown in Fig.~\ref{fig:2}, the prices generally increase with the residual load. 
This general dependency is well approximated by a linear function except for the extreme cases of very small residual load (i.e. a large portion of power being generated by renewable sources) or the opposite case of very large residual load (i.e. full conventional generation). 
We now examine in more detail the impact of large-scale weather regimes and systems on the statistics of electricity prices.

\begin{figure}[t]
	\includegraphics[width=\linewidth]{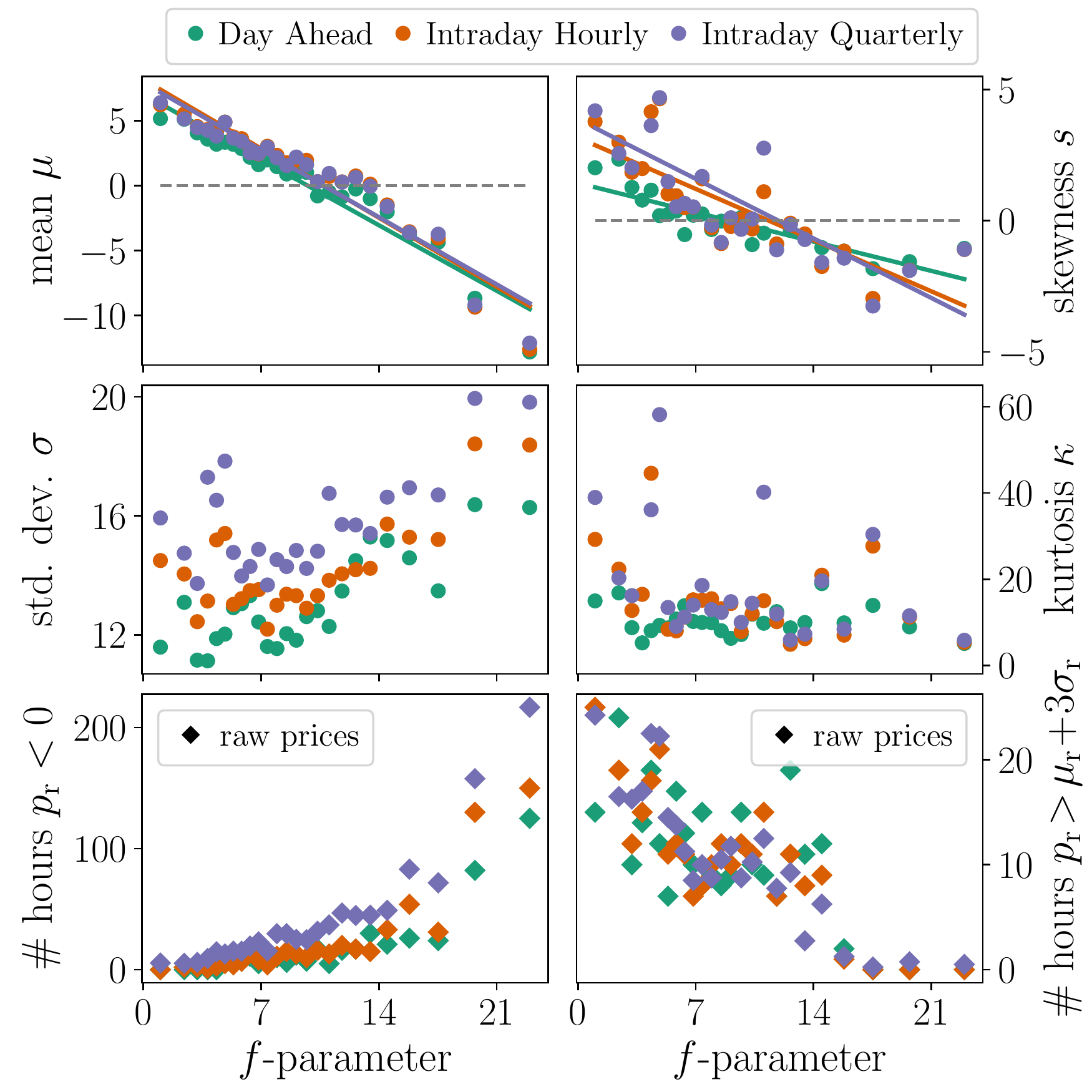}
	\caption{Impact of large-scale weather regimes on the statistics of electricity prices. We sort all time intervals according to the $f$-parameter, that represents the strength of the flow over Central Europe, and evaluate the statistics of the resulting subsets of the time series.
	We observe that low $f$-parameter values (calm wind conditions) are associated with a large mean $\mu$, small standard deviation $\sigma$, positive skewness $s$, and somewhat large kurtosis $\kappa$. 
	These are the weather periods with low renewable generation in Germany, which is dominated by wind generation. 
	When examining the raw prices (bottom plots), there are no periods of negative price (i.e. number of hours of negative prices), and a maximum of `high' price events (where the raw prices $p_{\mathrm{r}}>\mu_{\mathrm{r}}+ 3\sigma_{\mathrm{r}}$, ${\mathrm{r}}$ for raw).
	As the $f$-parameter value increases (windier conditions over Central Europe) the mean $\mu$ and skewness $s$ turns negative, the standard deviation $\sigma$ increases, and the kurtosis $\kappa$ decreases slightly.
	Likewise, negative (raw) price events increase and `high' (raw) prices vanish.
	The top four plots utilise the processed price data, the bottom plots use the actual raw prices, to best showcase the `true' negative price events.}\label{fig:8}
\end{figure}

An objective method to characterise the large-scale circulation in the lower atmosphere is the circulation weather type (CWT) approach~\cite{Jones1993}, which has turned out to be particularly suitable for wind energy applications in Central Europe~\cite{Reyers2015,Wohland2018,Weber2018,Weber2019}. 
In this approach mean sea level pressure (MSLP) fields around a central point in Central Europe (here: 10° East and 50° North near Frankfurt/Main, Germany) are assigned to one out of eight directional and/or two rotational weather types. 
Further, the strength of the flow is calculated and provided as the $f$-parameter. 
Low values of the $f$-parameter represent weak pressure gradients across Central Europe and are thus associated with weak winds, while high $f$-parameter values are related to strong pressure gradients and high wind speeds. 
In this study we use hourly MSLP fields of the latest reanalysis dataset of the European Centre for Medium-Range Weather Forecasts (ERA5~\cite{Hersbach2000}).

In Fig.~\ref{fig:8} the relationship between the hourly $f$-parameter and different statistics of the electricity price time series is shown. 
To this end, we condition the price time series to different intervals of the $f$-parameter and evaluate the statistical moments separately for each segment.
A distinct impact of the $f$-parameter is revealed for the mean $\mu$ and the skewness $s$, which are both positive for low $f$-parameter values and become negative for high values. 
This indicates that under calm wind conditions elevated average prices with a skewed distribution towards high price events occur. 
In contrast, high pressure gradients and the associated strong surface winds result in reduced average prices with a skewed distribution towards negative price events. This is valid for all three electricity markets. 
The standard deviation $\sigma$ (kurtosis $\kappa$) tends to increase (decrease) with increasing $f$-parameter, thus being characteristic of prices during periods of high renewable penetration, but the trends are less clear when compared to mean $\mu$ and the skewness $s$. 
The change of the mean $\mu$ from positive to negative values for increasing $f$-parameter agrees well with the merit-order effect, indicating a fundamental change in the shape of the distribution of prices as the weather changes.
Further, we found a maximum of `high' price events and a minimum of negative electricity prices during calm wind conditions ($f$-parameter almost zero), while negative price events increase and `high' prices vanish during windy periods (high $f$-parameter).

Interestingly, it is mainly the strength and not the direction of the flow that dominates the statistical moments of the electricity price time series. 
An agreeing yet small correlation in the statistics of prices were found when analysing the directional CWT West (associated with a strong zonal flow over Central Europe) and the rotational CWT Anticyclonic (associated with a stable high-pressure system) separately (see App.~\ref{app:2}).

Lastly, we are left with the question of whether intrinsic memory in price time series (as observed in Fig.~\ref{fig:6}) is in line with the intrinsic changes of the atmospheric flow over Central Europe.
In Fig.~\ref{fig:6} we show the auto-correlation of the $f$-parameter, which falls in line with the typical decay of the auto-correlation of price time series. 
Hence, the changes in large-scale weather regime may provide a physical justification for the adequacy of a superstatistical treatment and the typical time scales of synoptic circulation patterns like high and low pressure systems may explain the observed superstatistical time scale $T$.
We should nevertheless note that price time series statistics also changes due to the workweek-weekend changes in generation and consumption, which falls outside the scope of this article.

\section{Conclusion}\label{sec:conclusion}

In this article, we have examined three price time series from the German and Austrian electricity market, indexed in the European Power Exchange (EPEX SPOT), from 2015 to 2019.
We analysed spot market prices: the day-ahead hourly electricity price, the intraday hourly electricity price, and the intraday quarter-hourly electricity price.
We focused particularly on explaining and justifying the very heavy tails evidence in the distributions of all price time series.
The three examined price time series are intrinsically correlated as they reflect the trade of electricity futures in an open market, associated with an identical initial evaluation of electricity price in Europe.
The four central questions addressed in this article were: 
1) What is an adequate model to describe the leptokurtic distribution of the price time series? 
2) What characteristics and time scales of the data give rise to these distributions? 
3) Can we determine these time scales from the data and find a physical explanation for these?
4) How is the above related to weather changes?

To tackle the first question, we started by addressing the presence of strong non-stationarities in the data.
Upon examining these time series, one is immediately noticing the strong non-stationary effects.
This is evidenced across several time scales: the average annual price is different every year; it varies over the months, and is often higher at the last month of each year; it varies between weekdays and weekends, and day and night. 
To pursue a statistical examination of the prices, we proposed a simple, purely data-driven detrending of the data via the Hilbert--Huang transform, with which we subtracted the slower trends of the price time series.

We then turned to finding an adequate distribution for the price time series data.
After removing the long term non-stationarities, we presented two general distributions to describe the heavily leptokurtic distributions of the price time series: $q$-Gaussians and symmetric Lévy $\alpha$-stable distributions.
We evaluated the quality of the fits of these two distributions by examining the Kullback--Leibler divergence between the proposed distributions and the empirical distributions of the price time series.
We found that $q$-Gaussian distributions offer, for all time series, the best fit, and from these fits extracted their $q$ value (all roughly $q = 1.5$).
Large values of $q$ imply heavy tailed distributions, which are also observed in other financial market time series~\cite{Borland2002,Borland2002b,Beck2009,Zhao2018,SosaCorrea2018,AlonsoMarroquin2019}.

This led us to the second and third question, what are the intrinsic time scales of these time series?
We uncovered the correlations in the price time series, i.e. their persistence and long-range dependence.
We found that all three price time series are highly anti-correlated on time scales $>12$ hours, having a Hurst coefficient $H=0.1\sim0.2$.
Moreover, we also unveiled a small scale phenomenon for the period $<12$ hours, where prices in both hourly markets become positively correlated with $H\approx0.6$, whereas the quarter-hourly market remains anti-correlated.
We note here that these markets show strong anti-persistence in periods $>12$ hours, rendering it conceptually possible to hedge prices~\cite{Bessembinder2002}. 
Statistically speaking, this trend means if the price is decreasing at a given moment, the price is very likely to increase in the next moment and vice versa.
This is a first intrinsic time scale we extract from the price time series, which relates to internal correlations of prices.

Subsequently, we examined the multifractal characteristics of the data.
We found a clear separation of the multifractal spectrum, described by $\Delta \hat{\alpha} \neq 0$ for the different time series, resulting in large widths at small time scales.
This indicates that very large deviations happen and correct themselves in a very short manner, under $<12$ hours.
Moreover, at the intermediate scale, the largest market, the day-ahead hourly market, shows the smallest multifractal spectrum width $\Delta \hat{\alpha}\approx0.5$, pointing to a time scale that ``anchors'' the price fluctuation -- i.e. a scale where fast price changes and jumps are not seen.
This serves as an base for the price of the other markets and ensures that no extreme events extend beyond this period.
This agrees with the common understanding that electricity prices can see very sharp peaks in prices, but this behaviour is unsustainable for long periods of time -- i.e. any fast change to extreme prices is very quickly corrected.
This constitutes the second time scale, from $12$ to $48$ hours, where extreme prices can happen but are corrected.

We returned to the overarching question of non-stationarity in price time series, proposing to describe the price time series distribution via a superposition of symmetric yet unspecified simple distributions.
Using superstatistical analysis, we showed that by assuming the underlying fundamental distributions to be symmetric, one can uncover a unique long time scale -- the long superstatistical time -- at roughly $96$ hours, for all three markets.
This constitutes the third intrinsic time scale extracted in this article, and it relates to the slow-changing non-stationarity effects in price time series.
Having uncovered the large time scale of changes in price time series we returned to our initially proposed $q$-Gaussian distributions of price time series.
From the superstatistical analysis, we extracted the entropic indices $\bar{q}$ -- a measure of the ``changes of statistics'' -- of each price time series, all roughly $\bar{q}=1.5$, which agree well with the fits from the aforementioned $q$ values of the $q$-Gaussian fits.
Hence, we offered an explanation for the largely leptokurtic distributions of price time series as a combination of the changing local statistics.

As a final step with respect to question 4, we examine Circulation Weather Types, and in particular the `$f$-parameter', which is a measure for the strength of the large-scale near-surface flow over Central Europe.
We found that European electricity price time series are highly dependent on the strength of the flow (rather than on the direction of the flow).
In particular, wind energy generation -- which depends on the pressure gradient over Europe -- is the main renewable energy generation type in Germany, and thus highly influences electricity prices.
We observe a clear relation between strength of the flow and the change in price dynamics: `calm' wind conditions (low $f$-parameter values) lead to price distributions with higher mean and positive skewness (i.e. more high-price events).
Similarly, these show a lower standard deviation, characteristic of a reliance on conventional generation.
On the opposite spectrum, strong pressure gradients (high $f$-parameter values) with windier conditions lead to low prices on average, negative skewness, and increased standard deviation.
We also observe a congruent auto-correlation decay of the electricity price and the $f$-parameter, which strongly suggest that the vanishing memory in the prices being induced by a change in the weather conditions, as we have observed in the superstatistical analysis.
Hence, we found a possible physical mechanism that explains the long superstatistical time of $\sim96$ hours for the price time series.

The analysis presented in this article provides some powerful and novel tools for a better understanding of spot market electricity price time series, which we investigated in this study for data from Germany and Austria, from 2015 to 2019.
Particularly, our methods may help to pave the way forward to enable modelling price time series with the correct statistical properties in future studies, by considering relevant characteristics like non-stationarities, adequate local distributions, and intrinsic correlations.
Our methods may also help in extracting information on the relevant time scales of transitions in given data and clarify their relation to weather changes.

\begin{acknowledgements}
We are grateful for discussions with Benjamin Schäfer, Sonja Germscheid, and Eike Cramer.
P.C.B., D.W., and L.R.G. gratefully acknowledge support from the Helmholtz Association via the grant \textit{Uncertainty Quantification -- From Data to Reliable Knowledge (UQ)} with no.~ZT-I-0029.
C.H. and D.W. gratefully acknowledge support from the German Federal Ministry of Education and Research with grant no.~03EK3055B.
This work was performed by L.R.G. as part of the Helmholtz School for Data Science in Life, Earth and Energy (HDS-LEE).
\end{acknowledgements}

\appendix
\section{Distributions of the volatilities \texorpdfstring{$\beta$~}~of the price time series}\label{app:1}

\begin{figure}[b]
	\includegraphics[width=\linewidth]{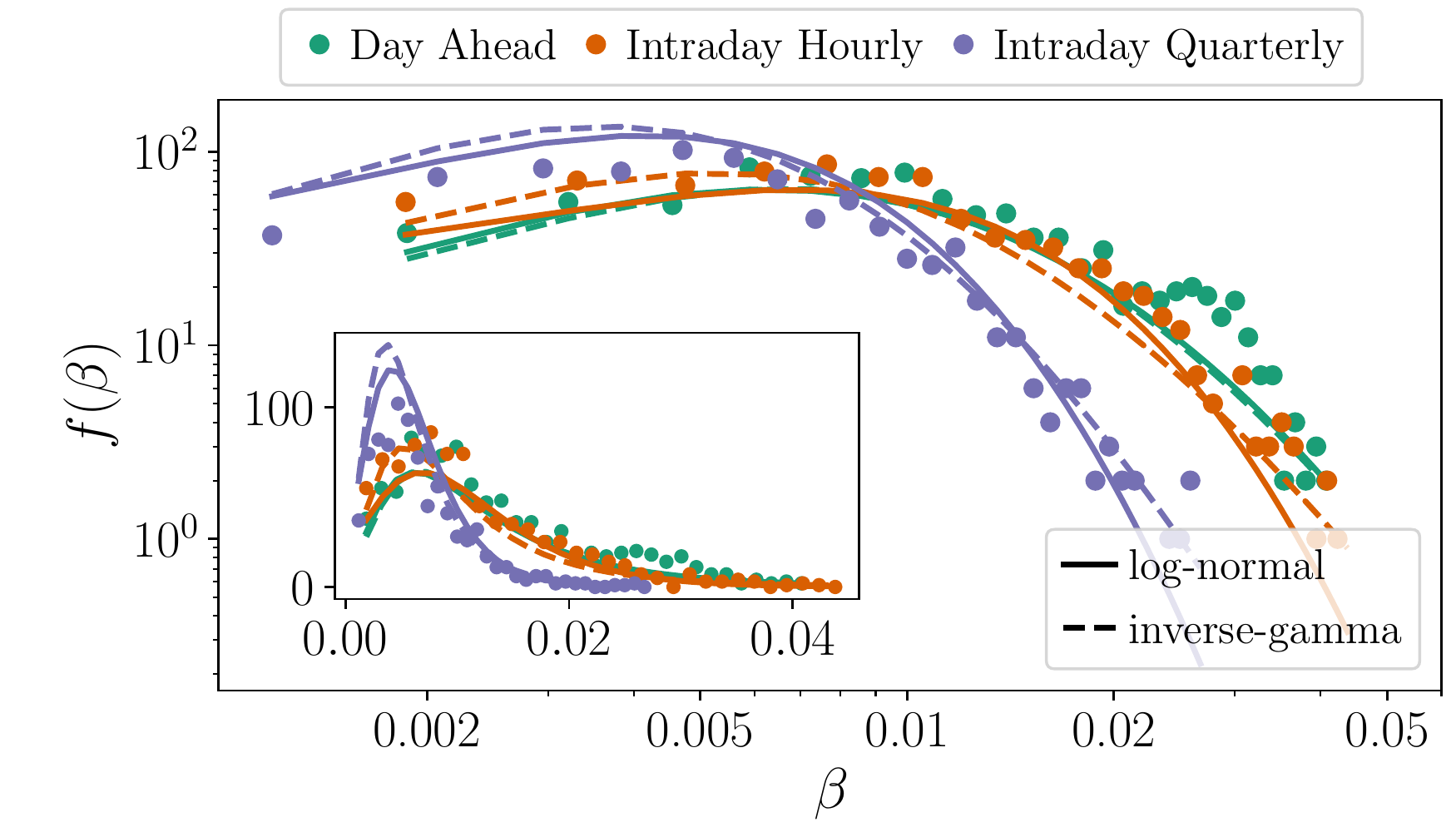}
	\caption{Distributions $f(\beta)$ of the volatilities $\beta$, in double logarithmic scale, of the three price time series and the two best-fitting log-normal and inverse-Gamma distributions, given in Eq.~\eqref{eq:dist_lognormal} and Eq.~\eqref{eq:dist_inverse_gamma} minimising the Kullback--Leibler divergence $D_{\mathrm{KL}}$ as given in Eq.~\eqref{eq:KL}.
	Inset shows a linear scale.}\label{fig:9}
\end{figure}

We present here the empirical distributions of the volatilities $\beta$ as drawn from Eq.~\eqref{eq:beta}, alongside with two candidate distributions, the log-normal distribution
\begin{equation}\label{eq:dist_lognormal}
    f_{\mathrm{log}\mathcal{N}}(\beta) = \frac{1}{\sqrt{2\pi}s\beta}\mathrm{exp}{\left(-\frac{\left(\ln{\beta}-\mu\right)^2}{2s^2}\right)},
\end{equation}
and the inverse-Gamma distribution
\begin{equation}\label{eq:dist_inverse_gamma}
    f_{\mathrm{inv}\Gamma}(\beta) = \frac{b^c}{\Gamma(c)}\frac{1}{\beta^{c+1}}\mathrm{exp}{\left(-\frac{b}{\beta}\right)}.
\end{equation}
In Fig.~\ref{fig:9} the empirical distributions and these two best-fitting distributions are shown.
We also tested fittings with Gamma and $F$-distributions, by minimising the Kullback--Leibler divergence $D_{\mathrm{KL}}$ as given in Eq.~\eqref{eq:KL}.
The log-normal and inverse-Gamma distribution provide the best fits.
These results must be judged as illustrative, as the data is insufficient in size to clearly single out a particular form of $f(\beta)$.
We can nevertheless see that the volatilites $\beta$ vary in a wide range of values, as described by an entropic indices $\bar{q}$ that deviates substantially from $1$.

\section{Dependence of price time series on atmospheric flow direction}\label{app:2}

\begin{figure}[b]
	\includegraphics[width=\linewidth]{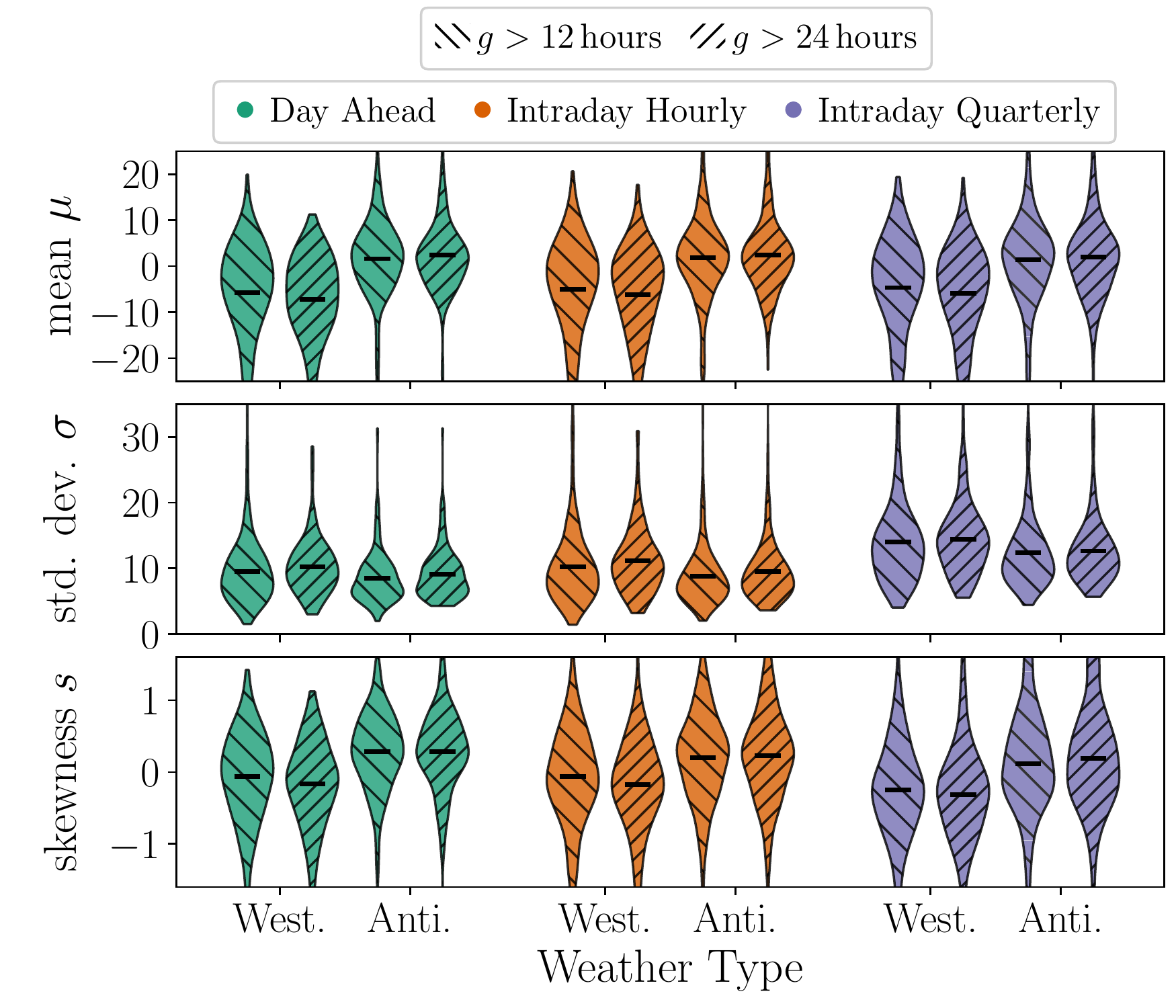}
	\caption{Impact of particular weather types on the statistics of price time series.
	We condition the price time series to either Westerly or Anticyclone states, for segments longer than 12 hours ($g>12$) and for segments longer than 24 hours ($g>24$).
	For each we calculate the mean $\mu$, standard deviation $\sigma$, and skewness $s$ of the prices.
	A weak separation in agreement with the price relation with the $f$-parameter given in Fig.~\ref{fig:8} can be seen.
	Westerly weather types are associated with a negative mean, more elevated standard deviation, and slightly negative kurtosis.
	In opposition, Anticyclone are associated with positive mean, smaller standard deviation, and slightly positive skewness.}\label{fig:10}
\end{figure}

The circulation weather typing approach, as discussed in Sec.~\ref{sec:cwt}, also enables a separation of the atmospheric flow into eight directional and/or two rotational types.
Following the work by Wohland~\textit{et~al.}~\cite{Wohland2018}, we focused on two of these weather types in this study: Anticyclonic and westerly weather types (note that for both weather types the full spectrum of potential $f$-parameter values is considered).
The Anticyclonic weather type is typically associated with stable and steady weather, while the westerly type often comes along with strong pressure gradients and thus strong winds, and with the passage of lows and fronts.
We conditioned the price time series to these two weather types.
To exclude situations where the atmospheric flow is strongly alternating on very short time scales, we select only the cases where the Anticyclonic and the westerly weather types prevail for longer than 12 hours and for longer than 24 hours, respectively.
Subsequently we analyse various statistics of prices for each separate segment of the price time series.
In Fig.~\ref{fig:10} we summarise the results, were we show the mean $\mu$, standard deviation $\sigma$, and skewness $s$ for each price time series, conditioned to either Westerly or Anticyclone states for longer than $12$ or $24 hours$, which we denote as $g>12\,$hours and $g>24\,$hours, respectively.
We do not include the kurtosis, as estimating the kurtosis requires a larger set of data points. 
In general, westerly weather types are rather associated with a negative mean, a slightly negative skewness, and a higher standard deviation when compared to the anticyclonic weather type.
These statistics reflect the typical characteristics of the westerly weather types with fluctuating strong wind speeds.
In contrast, the mean and the skewness tend to positive values for the anticyclonic weather types, and the standard deviation is smaller than for the westerly weather types.
These results generally agree with what we obtained for the $f$-parameter in Sec.~\ref{sec:cwt}, yet the effect is considerably smaller than for the $f$-parameter.

\bibstyle{apsrev4-2}
\bibliography{bib}

\end{document}